\documentstyle[psfig,12pt]{article}

\newcommand{\be}{\begin{equation}}
\newcommand{\ee}{\end{equation}}
\newcommand{\bea}{\begin{eqnarray}}
\newcommand{\eea}{\end{eqnarray}}
\newtheorem{lem}{Lemma}[section]
\newtheorem{pro}{Proposition}[section]

\newtheorem{defn}{Definition}[section]
\renewcommand{\thefootnote}{\alph{footnote}}

\def\bbbr{{\rm I\!R}} 

\def\bbbn{{\rm I\!N}} 

\def\bbbm{{\rm I\!M}}

\def\bbbh{{\rm I\!H}}

\def\bbbf{{\rm I\!F}}

\def\bbbk{{\rm I\!K}}

\def\bbbl{{\rm I\!L}}

\def\bbbp{{\rm I\!P}}

\def\bbbe{{\rm I\!E}}

\def\bbbone{{\mathchoice {\rm 1\mskip-4mu l} {\rm 1\mskip-4mu l}
{\rm 1\mskip-4.5mu l} {\rm 1\mskip-5mu l}}}

\def\bbbc{{\mathchoice {\setbox0=\hbox{$\displaystyle\rm C$}\hbox{\hbox
to0pt{\kern0.4\wd0\vrule height0.9\ht0\hss}\box0}}
{\setbox0=\hbox{$\textstyle\rm C$}\hbox{\hbox
to0pt{\kern0.4\wd0\vrule height0.9\ht0\hss}\box0}}
{\setbox0=\hbox{$\scriptstyle\rm C$}\hbox{\hbox
to0pt{\kern0.4\wd0\vrule height0.9\ht0\hss}\box0}}
{\setbox0=\hbox{$\scriptscriptstyle\rm C$}\hbox{\hbox
to0pt{\kern0.4\wd0\vrule height0.9\ht0\hss}\box0}}}}

\def\bbbq{{\mathchoice {\setbox0=\hbox{$\displaystyle\rm Q$}\hbox{\raise
0.15\ht0\hbox to0pt{\kern0.4\wd0\vrule height0.8\ht0\hss}\box0}}
{\setbox0=\hbox{$\textstyle\rm Q$}\hbox{\raise
0.15\ht0\hbox to0pt{\kern0.4\wd0\vrule height0.8\ht0\hss}\box0}}
{\setbox0=\hbox{$\scriptstyle\rm Q$}\hbox{\raise
0.15\ht0\hbox to0pt{\kern0.4\wd0\vrule height0.7\ht0\hss}\box0}}
{\setbox0=\hbox{$\scriptscriptstyle\rm Q$}\hbox{\raise
0.15\ht0\hbox to0pt{\kern0.4\wd0\vrule height0.7\ht0\hss}\box0}}}}
\def\bbbt{{\mathchoice {\setbox0=\hbox{$\displaystyle\rm
T$}\hbox{\hbox to0pt{\kern0.3\wd0\vrule height0.9\ht0\hss}\box0}}
{\setbox0=\hbox{$\textstyle\rm T$}\hbox{\hbox
to0pt{\kern0.3\wd0\vrule height0.9\ht0\hss}\box0}}
{\setbox0=\hbox{$\scriptstyle\rm T$}\hbox{\hbox
to0pt{\kern0.3\wd0\vrule height0.9\ht0\hss}\box0}}
{\setbox0=\hbox{$\scriptscriptstyle\rm T$}\hbox{\hbox
to0pt{\kern0.3\wd0\vrule height0.9\ht0\hss}\box0}}}}

\def\bbbs{{\mathchoice
{\setbox0=\hbox{$\displaystyle     \rm S$}\hbox{\raise0.5\ht0\hbox
to0pt{\kern0.35\wd0\vrule height0.45\ht0\hss}\hbox
to0pt{\kern0.55\wd0\vrule height0.5\ht0\hss}\box0}}
{\setbox0=\hbox{$\textstyle        \rm S$}\hbox{\raise0.5\ht0\hbox
to0pt{\kern0.35\wd0\vrule height0.45\ht0\hss}\hbox
to0pt{\kern0.55\wd0\vrule height0.5\ht0\hss}\box0}}
{\setbox0=\hbox{$\scriptstyle      \rm S$}\hbox{\raise0.5\ht0\hbox
to0pt{\kern0.35\wd0\vrule height0.45\ht0\hss}\raise0.05\ht0\hbox
to0pt{\kern0.5\wd0\vrule height0.45\ht0\hss}\box0}}
{\setbox0=\hbox{$\scriptscriptstyle\rm S$}\hbox{\raise0.5\ht0\hbox
to0pt{\kern0.4\wd0\vrule height0.45\ht0\hss}\raise0.05\ht0\hbox
to0pt{\kern0.55\wd0\vrule height0.45\ht0\hss}\box0}}}}

\def\bbbz{{\mathchoice {\hbox{$\sf\textstyle Z\kern-0.4em Z$}}
{\hbox{$\sf\textstyle Z\kern-0.4em Z$}}
{\hbox{$\sf\scriptstyle Z\kern-0.3em Z$}}
{\hbox{$\sf\scriptscriptstyle Z\kern-0.2em Z$}}}}

\def\bC{\ifmmode{\bbbc }\else${\bbbc }$\fi}
\def\bE{\ifmmode{\bbbe }\else${\bbbe }$\fi}
\def\bF{\ifmmode{\bbbf }\else${\bbbf }$\fi}
\def\bH{\ifmmode{\bbbh }\else${\bbbh }$\fi}
\def\bI{\ifmmode{\bbbone }\else${\bbbone }$\fi}
\def\bK{\ifmmode{\bbbk }\else${\bbbk }$\fi}
\def\bL{\ifmmode{\bbbl }\else${\bbbl }$\fi}
\def\bM{\ifmmode{\bbbm }\else${\bbbm }$\fi}
\def\bN{\ifmmode{\bbbn }\else${\bbbn }$\fi}
\def\bP{\ifmmode{\bbbp }\else${\bbbp }$\fi}
\def\bQ{\ifmmode{\bbbq }\else${\bbbq }$\fi}
\def\bR{\ifmmode{\bbbr }\else${\bbbr }$\fi}
\def\bS{\ifmmode{\bbbs }\else${\bbbs }$\fi}
\def\bZ{\ifmmode{\bbbz }\else${\bbbz }$\fi}

\def\cA{\ifmmode{\cal A}\else${\cal A}$\fi}
\def\cB{\ifmmode{\cal B}\else${\cal B}$\fi}
\def\cC{\ifmmode{\cal C}\else${\cal C}$\fi}
\def\cD{\ifmmode{\cal D}\else${\cal D}$\fi}
\def\cE{\ifmmode{\cal E}\else${\cal E}$\fi}
\def\cF{\ifmmode{\cal F}\else${\cal F}$\fi}
\def\cG{\ifmmode{\cal G}\else${\cal G}$\fi}
\def\cH{\ifmmode{\cal H}\else${\cal H}$\fi}
\def\cI{\ifmmode{\cal I}\else${\cal I}$\fi}
\def\cJ{\ifmmode{\cal J}\else${\cal J}$\fi}
\def\cK{\ifmmode{\cal K}\else${\cal K}$\fi}
\def\cL{\ifmmode{\cal L}\else${\cal L}$\fi}
\def\cM{\ifmmode{\cal M}\else${\cal M}$\fi}
\def\cN{\ifmmode{\cal N}\else${\cal N}$\fi}
\def\cO{\ifmmode{\cal O}\else${\cal O}$\fi}
\def\cP{\ifmmode{\cal P}\else${\cal P}$\fi}
\def\cQ{\ifmmode{\cal Q}\else${\cal Q}$\fi}
\def\cR{\ifmmode{\cal R}\else${\cal R}$\fi}
\def\cS{\ifmmode{\cal S}\else${\cal S}$\fi}
\def\cT{\ifmmode{\cal T}\else${\cal T}$\fi}
\def\cU{\ifmmode{\cal U}\else${\cal U}$\fi}
\def\cV{\ifmmode{\cal V}\else${\cal V}$\fi}
\def\cW{\ifmmode{\cal W}\else${\cal W}$\fi}
\def\cX{\ifmmode{\cal X}\else${\cal X}$\fi}
\def\cY{\ifmmode{\cal Y}\else${\cal Y}$\fi}
\def\cZ{\ifmmode{\cal Z}\else${\cal Z}$\fi}

\def\bk#1{{\ifmmode{\langle#1\rangle}\else${\langle#1\rangle}$\fi}} 

\def\sp#1#2{{\ifmmode{\langle{#1}|{#2}\rangle}
\else${\langle{#1}|{#2}\rangle}$\fi}}

\def\prf{\medbreak\noindent{\bf Proof.}\enspace}

\def\qed{\hfill $\sqcap$\llap{$\sqcup$}\break}

\def\data{\the\day\space\ifcase\month\or January \or February \or March \or
April \or May \or June \or July \or August \or September
\or October \or November \or December \fi\space\the\year}

\def\date{\the\day\space\ifcase\month\or Janvier \or F\'evrier \or Mars \or
Avril \or Mai \or Juin \or Juillet \or Ao\^ut \or Septembre
\or Octobre \or Novembre \or D\'ecembre \fi\space\the\year}

\def\sc{\scriptstyle}

\def\vec#1{\ifmmode
\mathchoice{\mbox{\boldmath$\displaystyle\bf#1$}}
{\mbox{\boldmath$\textstyle\bf#1$}}
{\mbox{\boldmath$\scriptstyle\bf#1$}}
{\mbox{\boldmath$\scriptscriptstyle\bf#1$}}\else
{\mbox{\boldmath$\bf#1$}}\fi}

\def\math#1{\ifmmode
\mathchoice{\mbox{$\displaystyle\rm#1$}}
{\mbox{$\textstyle\rm#1$}}
{\mbox{$\scriptstyle\rm#1$}}
{\mbox{$\scriptscriptstyle\rm#1$}}\else
{\mbox{$\rm#1$}}\fi}

\def\accol2#1#2{\left\{ 
\begin{array}{l}#1\\#2\end{array}
\right.}					
\def\matrix22#1#2#3#4{\left(
\begin{array}{cc}#1&#2\\#3&#4\end{array}
\right)}					

\def\e{\math{e}}				

\def\ds{\displaystyle}
\def\psdir{.}
\def\eset{\emptyset}
\def\nnb#1#2{<\!#1,#2\!>}
\def\s{\sigma}
\def\t{\tau}
\def\Js{J_\sigma}
\def\Jt{J_\tau}
\def\Jst{J_{\sigma\tau}}
\def\w{\omega}

\def\th{\math{tanh}}
\def\ch{\math{cosh}}
\def\sh{\math{sinh}}
\def\ash{\math{argsinh}}
\def\req#1{(\ref{#1})}
\def\isdefby{:=}
\def\defines{=:}
\def\Zpp{{\cZ^{++}}}
\def\Zppnn{{\Xi^{(+,+)}}}
\def\Zf{{\cZ^f}}
\def\Zfnn{{\Xi^f}}

\def\Zrc*{{\cZ_{RC}^{*}}}

\def\ns{n_\s}
\def\nt{n_\t}
\def\nsb{n_{\s,b}}
\def\ntb{n_{\t,b}}
\def\uns{{\underline n}_\s}
\def\unt{{\underline n}_\t}
\def\un{{\underline{n}}}

\def\hn{\hat {n}}
\def\uhns{\hat {\underline n}_\s}
\def\uhnt{\hat {\underline n}_\t}
\def\uhn{\hat{\underline n}}
\def\gs{\gamma_\s}
\def\gt{\gamma_\t}
\def\ha{\hat{a}}
\def\hq{\hat{q}}
\def\hb{{\hat{b}}}
\def\a0{\alpha_0}
\def\as{\alpha_\s}
\def\at{\alpha_\t}
\def\ast{\alpha_{\s\t}}
\def\ugs{{\underline \gamma}_\s}
\def\ugt{{\underline \gamma}_\t}
\def\phm{\phantom{m}}
\def\nnm{\nonumber}

\def\conn{\mathop\leftrightarrow}
\def\sconn{\mathop\leftrightarrow^\s}

\def\rem{{\bf Remark: }}
\def\ex{{\bf Example:}}
\def\Ns{{N_\s}}
\def\Nt{{N_\t}}
\def\mupp{\mu_{\Lambda}^{++}}
\def\mupm{\mu_{\Lambda}^{+-}}
\def\mump{\mu_{\Lambda}^{-+}}
\def\mumm{\mu_{\Lambda}^{--}}
\def\muf{\mu_{\Lambda}^f}
\def\muAT#1#2{\mu_\Lambda^{#1}(#2)}
\def\muATth#1#2{\mu^{#1}(#2)}
\def\murc{\rho_\Lambda}
\def\mugrcp{\nu_\Lambda^+}
\def\mugrcf{\nu_\Lambda^f}
\def\mugrc{\nu_\Lambda}
\def\hmugrc{\widehat{\nu}_\Lambda}
\def\mugrcb{\nu}
\def\mup{\zeta}
\def\mugp{\lambda}
\def\muIs{\eta_\Lambda}

\def\onsb{{\overline n}_{\s,b}}
\def\ontb{{\overline n}_{\t,b}}
\def\on{{\overline n}}
\def\om{\omega}
\def\defsty{\bf}

\def\jref#1#2#3#4#5{#1 {\bf #2} (19#3), #4--#5.}

\newcommand{\mybibitem}[9]{\bibitem[#1]{#2}#3, {\sl #4}, \jref{#5}{#6}{#7}{#8}{#9}}
\newcommand{\mybibprep}[4]{\bibitem[#1]{#2}#3, Preprint, #4}
\newcommand{\mybibbook}[5]{\bibitem[#1]{#2}#3, {\sl #4}, {#5}}

\def\cmp{Commun. Math. Phys.}
\def\jsp{J. Stat. Phys. }

\def\jpC{J. Phys. C: Solid State Phys.}

\def\pr{Phys. Rev.}
\def\prB{Phys. Rev. B}
\def\prE{Phys. Rev. E}

\def\ptrf{Probab. Theory Relat. Fields}
\def\phy{Physica}

\begin{document}
\title{Random--Cluster Representation of the Ashkin--Teller Model}
\author{C.-E. Pfister$^1$ and Y. Velenik$^2${$\,^i$}\\\\
$^1$ D\'epartement de Math\'ematiques, E.P.F.-L.,\\CH--1015 Lausanne, Switzerland\\
$^2$ Institut de Physique Th\'eorique, E.P.F.-L.,\\CH--1015 Lausanne, Switzerland\\}
\maketitle
\vspace*{-10.2cm}{\bf JSP 96-233}\vspace*{10.2cm}
\renewcommand{\thefootnote}{\roman{footnote}}
\footnotetext[1]{Supported by Fonds National Suisse
Grant 2000-041806.94/1}
\renewcommand{\thefootnote}{\arabic{footnote}}
\begin{abstract}
We show that a class of spin models, containing the Ashkin--Teller model, admits a generalized
random--cluster (GRC) representation. Moreover we show that basic properties of the usual
representation,
such as FKG inequalities and comparison inequalities, still hold for this generalized
random--cluster model. Some elementary consequences are given. We
also consider the duality transformations in the spin representation and in the GRC model and
show that they commute.
\end{abstract}

The introduction by Fortuin and Kasteleyn \cite{FoKa,Fo1,Fo2} of the random--cluster model
in the late 60s has given
rise to numerous important results. First it provided a unified representation of
several famous models, including the Ising, Potts and percolation models, thus allowing the
comparison between them. It also brought a whole class of models interpolating between
the latter ones. The random--cluster representation has been used in many recent proofs in
statistical mechanics, for example in large deviations theory \cite{Io,Pi}. The fact is that this
model has
several nice properties, as FKG and comparison inequalities, allowing to derive 
non--perturbative results for the original models. One of the properties which has also often
been used is that the two-dimensional random--cluster model is self--dual, and that this duality commutes
with the duality of the original models; this has been used for example in the study of the decay
of the connectivity in the Ising model \cite{ChChSc}. Other applications of this representation
have been found in numerical studies, in particular the Swendsen-Wang algorithm is based
on it.\\
It would then be interesting to be able to extend this representation to a wider class of models,
while keeping most of its properties. This appears to be possible. We show that the
Ashkin--Teller model (and a class of models generalizing the Ashkin--Teller model,
and containing the partially-symmetric Potts models) admits a similar representation,
which in fact generalizes the usual one. The nice point is that it is still possible to prove FKG
inequalities, comparison inequalities and commutativity of the dual transformations
 for this new representation.\\
Such a representation has already been considered in \cite{WiDo,SaSo}. The main
goal in these papers is to develop a Swendsen-Wang type algorithm for the
Ashkin-Teller model. A closely related representation has also
appeared in the study of partially symmetric Potts models \cite{LaMaRu}. 
Their representation appears as a
special case of the one studied here. Nevertheless, properties of the measure
were not studied in these papers.\\
Although the Ashkin--Teller model has been introduced more than half a century
ago
\cite{AsTe}, there are still several open questions about this model. Some of
the
tools developed for the study of the Potts model via the random--cluster
representation are
useful in the study of the Ashkin--Teller model. In this paper, we focus on the
properties of the
two-dimensional model, and give only some elementary applications of the
inequalities.
At the end of the paper, we discuss possible extensions of the results. We
shall consider more
elaborate applications in a separate publication.
One of the main points of the paper is to show that 
elementary methods can be used to study 
the duality transformation
of the spin model and the random-cluster representation. It is advantageous to 
derive the duality transformation using the high--temperature expansion based
on the elementary formula (\ref{straight1}); moreover, this approach allows to
study correlation functions and boundary conditions very explicitly. The
random-cluster representation is not more difficult than the high-temperature
expansion; it is based on the elementary formula (\ref{straight2}).

After we finished this work we received the paper \cite{CM} by L. Chayes and
J. Machta. In this paper graphical representations are developed for a variety of
spin-systems including the Ashkin-Teller model. These representations are used
in connection with Swendsen-Wang type algorithms. The case of the Ashkin-Teller
model is studied  in details. Although the presentation of the model is
different (compare e.g. the phase diagrams), essentially all our results about
the random-cluster model are explicitly derived in \cite{CM} (see in particular
Propositions 3.5 and 3.6 therein).

{\bf Acknowledgements:} We thank L.Chayes for discussions and communicating us
his results with Machta. We also acknowledge discussions with L.Laanait
and J.Ruiz about the duality transformation.
\section{The Ashkin--Teller model}
\label{sec_ATmodel}
\setcounter{equation}{0}
{\bf Lattices and cell-complexes}

The model is defined on $\bbbz^2$ or on some bounded subset $\Lambda \subset \bbbz^2$,
\be
\bbbz^2 \isdefby \{t=(t_1,t_2) : t_i \in \bbbz, \; i=1,2\}\,.
\ee
We call {\defsty sites $t$} the elements of the lattice $\bbbz^2$. Two sites $t$ and $t'$ are {\defsty 
nearest--neighbours} if $|t_1-t'_1|+|t_2-t'_2| = 1$. By definition the boundary of a site is
the empty set. We call {\defsty bonds $b=\nnb{t}{t'}$}
the subsets of 
$\bbbr^2$ which are straight line segments with the nearest--neighbours sites $t$ and $t'$
as endpoints. The boundary of a bond is $\delta b=\{t,t'\}$, and the boundary of a set \cB\/ of
bonds
is the set $\delta\cB=\{t \in \bbbz^2 : t \in \delta b$ for an odd number of bonds $b \in \cB\}$.
Finally
we call {\defsty plaquettes $p$} the subsets of $\bbbr^2$ which are unit squares whose corners are
sites. Their boundary is the set of the four bonds forming their boundary as a subset of $\bbbr^2$.
With this structure the lattice becomes a cell--complex, which we denote by $\bbbl$.

Another lattice is important, the dual lattice $(\bbbz^2)^*$,
 \be
(\bbbz^2)^* \isdefby \{t=(t_1,t_2) : t_i+1/2 \in \bbbz, \; i=1,2\}\;.
\ee
We can of course define the same objects as before for the dual lattice, they will be denoted
$t^*$, $b^*$ and $p^*$ respectively. The dual cell--complex will be denoted by $\bbbl^*$.
The following important geometrical relations hold:
\begin{enumerate}
\item each site $t$ is the center of a unique plaquette $p^*$,
\item each bond $b$ is crossed by a unique bond $b^*$,
\item each plaquette $p$ has a unique site $t^*$ at its center.
\end{enumerate}

A subset $\Lambda \subset \bbbz^2$ is {\defsty simply connected} if the subset of $\bbbr^2$ which
is the union of all plaquettes $p^*(t)$, $t \in \Lambda$, is a simply connected set in $\bbbr^2$.

{\bf Dual of a set}

Let $\Lambda \subset \bbbz^2$; we will also denote by $\Lambda$ the following subset of
$\bbbl$: the sites of $\Lambda$ are the elements of $\Lambda$ (as subset of $\bbbz^2$); the
bonds of $\Lambda$ are the bonds of $\bbbl$ whose boundary belongs to $\Lambda$; the
plaquettes of $\Lambda$ are the plaquettes $p$ of $\bbbl$ whose boundary is given by
four bonds of $\Lambda$. We will denote by $\cB(\Lambda)$ the set of bonds of $\Lambda$.\\
We now define a dual set for $\Lambda$. We will define another notion of dual set later (see
subsection \ref{geom_res}).\\
We define $\Lambda^* \subset
\bbbl^*$ in the following way: the plaquettes of $\Lambda^*$ are all plaquettes
of $\bbbl^*$ whose center is some site of $\Lambda$; the bonds of $\Lambda^*$ are all
bonds of $\bbbl^*$ belonging to the boundary of some plaquette in $\Lambda^*$; the sites
of $\Lambda^*$ are all sites of $\bbbl^*$ belonging to the boundary of some bonds in
$\Lambda^*$.\\

{\bf Configurations, Hamiltonian and Gibbs states}

A {configuration $\om$} of the model is an element of the product space
\be		
\Omega \isdefby [\{-1,1\}\times\{-1,1\}]^{\bbbz^2}\,.
\ee

The value of the configuration $\omega=(\s,\t)$ at $t\in\bbbz^2$ is $\omega(t)=(\sigma(t),\tau(t))$.\\
Let $\Lambda\subset \bbbz^2$. A configuration $\omega$ is said to satisfy the {\defsty
$(+,+)$-boundary condition} in $\Lambda$ if
\be
\omega(t) = (1,1)\;\;\;\forall t \notin \Lambda\,.
\ee
The Ashkin--Teller Hamiltonian on $\Lambda$ is
\be
H_\Lambda = - \sum_{\nnb{i}{j}: \atop \{i,j\}\cap \Lambda\neq\eset}\left\{\Js \s_i\s_j +
\Jt \t_i\t_j + \Jst \s_i\s_j\t_i\t_j\right\}\,,
\label{hamAT}
\ee
where $\Js$, $\Jt$ and $\Jst$ are real numbers called {\defsty coupling constants}.

The {\defsty Gibbs measure on $\Lambda$ with (+,+)-boundary condition} is the probability measure
given by the formula
\be\label{+bd}
\mupp(\omega)\isdefby \cases{
\Zppnn(\Lambda)^{-1}
\exp (-H_{\Lambda}(\omega))&  if $\omega(t)$ satisfies
the $(+,+)$-b.c. on $\Lambda$,\cr
0& otherwise.\cr}
\ee
where the normalization $\Zppnn(\Lambda)$ is called the {\defsty partition function with
$(+,+)$-boundary condition}.

In the same way, we can introduce {\defsty $(+,-)$-, $(-,+)$- and $(-,-)$-boundary conditions} by
imposing the corresponding value to $\omega$ outside $\Lambda$.\\
Notice that the Ashkin--Teller model has the following symmetries :
\be
\mupp((\s,\t)) = \mupm((\s,-\t)) = \mump((-\s,\t)) = \mumm((-\s,-\t))\,,
\ee
so we consider only $(+,+)$-boundary condition.\\

We also define the {\defsty Gibbs measure on $\Lambda$ with free boundary condition}
\be
\muf(\omega) \isdefby \Zfnn(\Lambda)^{-1}\prod_{\nnb{i}{j}\subset\Lambda}\exp\{\Js\s_i\s_j+\Jt\t_i\t_j
+\Jst \s_i\s_j\t_i\t_j\}\,,
\ee
where the normalization $\Zfnn(\Lambda)$ is called the {\defsty partition function with
free boundary condition}.\\

\rem For $\Jst = 0$, the Ashkin--Teller model reduces to 2 independent Ising models, while for
$\Js=\Jt=\Jst$ it becomes the 4-states Potts model.\\

We will always suppose that the coupling constants satisfy
\be
\Js \geq \Jt \geq \Jst\,.
\ee
Note that there is no loss of generality in doing this choice. Indeed we can always transform
\req{hamAT} to obtain this order.
For example, if $\Jst > \Jt$, then we can make the following change of variables:
$(\s_i,\t_i) \mapsto (\s_i,\theta_i)$, where $\theta_i = \s_i\t_i$.\\

In this paper, we further impose that
\be
\Js \geq 0,\phm \Jt \geq 0,\phm \th\Jst \geq -\th\Js\th\Jt\,.
\label{CondJsJtJst}
\ee
\section{Duality of the Ashkin--Teller model}
\setcounter{equation}{0}
Duality of the Ashkin--Teller model has been known for a long time \cite{Fa,Ba}. However,
for the sake of completeness, as well as to fix the notations which will be used when considering
the duality of the random--cluster model, we give here a straightforward account of this transformation.
\subsection{Low temperature expansion for $(+,+)$-boundary conditions}
\label{lowT}
Let $\Lambda \subset \bbbz^2$ be bounded and simply connected.Let us now consider the
Ashkin--Teller model defined on $\Lambda$, with
$(+,+)$-boundary conditions and with coupling constants $\Js$, $\Jt$ and $\Jst$.\\
With this kind of boundary conditions, we can describe geometrically all configurations
$(\s,\t)$ of the model by giving the sets
\bea
&{\cM}_\s \isdefby \{p^*(t) : t \in \Lambda,\s_t = -1\},&\nnm\\
&{\cM}_\t \isdefby \{p^*(t) : t \in \Lambda,\t_t = -1\}.&\nnm
\eea
The boundaries of these sets, considered as subsets of $\bbbr^2$, define two
sets of bonds of $\bbbl^*$. Maximal connected components $\gs$, $\gt$ of these
sets of bonds are called {\defsty $\s$- and $\t$-contours} respectively. We
will call {\defsty closed contours} contours  such that $\delta \gamma =
\eset$. The {\defsty length} of a contour is its cardinality as a set of bonds
and is denoted by $|\gamma|$.  A {\defsty configuration of contours} is a set
of closed contours such that: (a) any two $\s$-contours are disjoint (as sets of bonds
{\em and} sites); (b) any two $\t$-contours are disjoint (in the same sense).
(There is no constraint between the $\s$- and $\t$-contours.) Such a set will
be denoted by $(\ugs,\ugt)$, where $\ugs$ denotes the set of $\s$-contours and
$\ugt$ the set of $\t$-contours.
To each spin configuration $\omega = (\s,\t)$, it is possible to associate a
unique configuration $(\ugs,\ugt)$ of contours.

\rem If $\Lambda$ is simply connected, then the converse is also true. If it is
not simply connected, then it will generally be false. Indeed, suppose
$\Lambda$ is a square with some hole in it, with $(+,+)$-boundary condition.
Then only configurations of contours such that there is an even number of $\s$
(and $\t$) -contours winding around the hole correspond to some spin
configurations. This will be important when considering duality.

Let us now introduce the {\defsty weights} of contours
\bea
&\w_\s(\gs) \isdefby \exp(-2(\Js+\Jst) |\gs|)\,,\;\;\;\w_\s(\ugs) \isdefby
\prod_{\gs\in\ugs}\w_\s(\gs)\,,&\nnm\\
&\w_\t(\gt) \isdefby \exp(-2(\Jt+\Jst) |\gt|)\,,\;\;\;\w_\t(\ugt) \isdefby
\prod_{\gt\in\ugt}\w_\t(\gt)\,.&
\eea
Introducing the following interaction between the contours,
\be
\w_{\s\t}(\gs,\,\gt) \isdefby \exp(4\Jst|\ugs \cap \ugt|)\,,
\ee
where $|\ugs \cap \ugt|$ is the cardinality of the set of bonds belonging
simultaneously to $\ugs$ and $\ugt$, the partition function in $\Lambda$ with
$(+,+)$-boundary condition can be written
\be
\Zppnn_\Lambda = C_1\sum_{\ugs,\ugt}\w_\s(\ugs)\w_\t(\ugt)\w_{\s\t}(\ugs,\ugt)\,,
\ee
where $C_1$ is some constant depending on $\Lambda$ but not on the
configurations which does not affect the results below. The sum is over families
of closed $\s$- and $\t$-contours.

\rem If $\Jst>0$ the interaction is such that the $\s$- and $\t$-contours will
attract each other while they will repel each other when $\Jst<0$.

It is therefore natural to use a normalized partition function with $(+,+)$-boundary condition
which is defined as
\be
\Zpp_\Lambda \isdefby \sum_{\ugs,\ugt}\w_\s(\ugs)\w_\t(\ugt)\w_{\s\t}(\ugs,\ugt)\,.
\label{Zpp}
\ee
\subsection{High-temperature expansion for free boundary conditions}
Suppose $\Lambda \subset \bbbz^2$ is bounded and simply connected. Let $\Lambda^*$ be the
dual of $\Lambda$ as defined earlier. We consider the Ashkin--Teller
Hamiltonian on $\Lambda^*$ with free boundary condition and coupling constants $\Js^*$,
$\Jt^*$ and $\Jst^*$.

We now proceed in doing a high-temperature expansion of $\Zfnn$
\bea
\Zfnn &=& \sum_{\s,\t}\prod_{\nnb{i}{j}\subset \Lambda^*}(\ch\Js^* + \s_i\s_j \sh\Js^*)(\ch\Jt^* +
\t_i\t_j \sh\Jt^*)\times\nnm\\
&\phantom{=}&\hspace{6cm}\times(\ch\Jst^* + \s_i\s_j\t_i\t_j \sh\Jst^*)\nonumber\\
&=& (\ch\Js^*\ch\Jt^*\ch\Jst^*)^{|\cB(\Lambda)|}  \sum_{\s,\t}\prod_{\nnb{i}{j}}
(1 + \s_i\s_j \th\Js^*)\times\nnm\\
&\phantom{=}&\hspace{3.5cm}\times(1 + \t_i\t_j \th\Jt^*)(1 + \s_i\s_j\t_i\t_j \th\Jst^*)\,.
\label{HTexp}\label{straight1}
\eea
Defining 
\be\label{stl}
s=\th\Js^*,\;\; t=\th\Jt^*,\;\; l=\th\Jst^*\,,
\ee
and
\be
S = \frac{s+tl}{1+stl}\,,\; T = \frac{t+sl}{1+stl}\,,\; L = \frac{l+st}{1+stl}\,,
\label{STL}
\ee
the above sum becomes
\be
(1+stl)^{|\cB(\Lambda)|} \sum_{\s,\t}\prod_{\nnb{i}{j}}\left\{1+S \s_i\s_j + T \t_i\t_j + L\s_i\s_j\t_i\t_j\right\}\,.
\label{avexp}
\ee
Expanding the product, we obtain a sum of terms that can be indexed by $(\eta_\s,\eta_\t)$,
$\eta_\s$, $\eta_\t \in \{0,1\}^{\cB(\Lambda)}$ (we recall that $\cB(\Lambda)$ is the set of
bonds of the cell--complex $\Lambda$). This is done in the following way:
\begin{enumerate}
\item Each time we take one term $1$ in \req{avexp}, we set $$\eta_\s(\nnb{i}{j}) = 0,\;
\eta_\t(\nnb{i}{j}) = 0,$$
\item Each time we take one term $S\s_i\s_j$ in \req{avexp}, we set $$\eta_\s(\nnb{i}{j}) = 0,\;
\eta_\t(\nnb{i}{j}) = 1,$$
\item Each time we take one term $T\s_i\s_j$ in \req{avexp}, we set $$\eta_\s(\nnb{i}{j}) = 1,\;
\eta_\t(\nnb{i}{j}) = 0,$$
\item Each time we take one term $L\s_i\s_j\t_i\t_j$ in \req{avexp}, we set
$$\eta_\s(\nnb{i}{j}) = 1,\; \eta_\t(\nnb{i}{j}) = 1.$$
\end{enumerate}
To each of these pairs $(\eta_\s,\eta_\t)$ we associate a configuration of
$\s$- and $\t$-contours $(\ugs,\ugt)$, where the $\gs$ are maximal connected
components of $\{b \in \cB(\Lambda) : \eta_\s(b)=1\}$ and $\gt$ are maximal connected components of $\{b \in \cB(\Lambda) : \eta_\t(b)=1\}$.\\
Note that we have interchanged $\s$ and $\t$, for later convenience, see section \ref{sec_dual}.\\
We now sum over $\s$, $\t$. Using the fact that $\sum_\s\s_i^{2k+1}=
\sum_\t\t_i^{2k+1}=0$, $\forall k \in \bbbn$, we see that the only contributing configurations
are those with only closed $\s$- and $\t$-contours. We obtain
\be
\Zfnn_{\Lambda^*} =
(1+stl)^{|\cB(\Lambda)|}4^{|\Lambda|}\sum_{\ugs,\ugt}\left\{S^{|\ugt|-|\ugs \cap \ugt|}
T^{|\ugs|-|\ugs \cap \ugt|}L^{|\ugs \cap \ugt|}\right\}\,,
\ee
where $|\gamma|$ denotes the cardinal of $\gamma$,
considered as a set of bonds.
We define the normalized partition function with free boundary conditions to be
\be
\Zf_{\Lambda^*} \isdefby \sum_{\ugs,\ugt}\left\{{S^{|\ugt|-|\ugs \cap
\ugt|}}T^{|\ugs|-|\ugs \cap \ugt|}
L^{|\ugs \cap \ugt|}\right\}\,.
\label{Zf}
\ee
\subsection{Duality}
\label{sec_dual}
\begin{pro}\label{Prop_dual}
Let $\Lambda$ be a simply connected bounded subset of $\bbbz^2$. Let $\cD=\{(x,y,z)\in\bbbr^3
: x\geq y\geq z, y>0, \th z > -\th x\th y\}$. Let $(\Js,\Jt,\Jst)\in \cD$ be the coupling constants of
the Ashkin--Teller model defined on $\Lambda$ with
$(+,+)$-boundary conditions.
Then the following relations
\bea
S(\Js^*,\Jt^*,\Jst^*) &=& \exp(-2(\Jt+\Jst))\,,\nonumber\\
T(\Js^*,\Jt^*,\Jst^*) &=& \exp(-2(\Js+\Jst))\label{DualRel}\,,\\
L(\Js^*,\Jt^*,\Jst^*) &=& \exp(-2(\Js+\Jt))\,,\nonumber
\eea
(where $S$, $T$ and $L$ have been introduced in \req{STL}) define a bijection
from $\cD$ on itself, such that
\be
\Zpp_\Lambda(\Js,\Jt,\Jst) = \Zf_{\Lambda^*}(\Js^*,\Jt^*,\Jst^*)\,.
\ee
On the closure of $\cD$, the application is still well-defined, but takes values in ${\overline
\bbbr^3}$ and is no more everywhere invertible.
\end{pro}
The proof is straightforward algebra; it is given in the appendix.
\subsection{The self--dual manifold}
\begin{pro}
The self-dual manifold, i.e. the set of fixed points of the duality relations \req{DualRel},
is given by
\be
l = \frac{1-st-s-t}{1-st+s+t}\,,
\ee
where $s=\th \Js$, $t=\th \Jt$ and $l=\th \Jst$.
\end{pro}
\prf
We want to find the values of $\Js$, $\Jt$ and $\Jst$ such that $\Js^*=\Js$, $\Jt^*=\Jt$ and
$\Jst^*=\Jst$. In particular one must have
\be\label{selfdualrel}
\frac{l+ts}{1+stl}=\e^{-2(\Js+\Jt)} = \e^{-2(\Js^*+\Jt^*)} =
\frac{(1-s)(1-t)}{(1+s)(1+t)}\,.\nnm
\ee
We have used
\be
\e^{-2(x+y)}=\frac{(1-\th x)(1-\th y)}{(1+\th x)(1+\th y)}\,.
\ee
After some algebraic manipulations, \req{selfdualrel} can be seen to be equivalent to
\be
l = \frac{1-st-s-t}{1-st+s+t}\,.
\ee
The two other relations are seen to be satisfied for these values of $l$ by substitution,
\bea
\frac{s+tl}{1+stl} = \frac{(1-t)(s+t)}{(1+t)(1-st)} = \frac{(1-t)(1-l)}{(1+t)(1+l)}\,,\nnm\\
\frac{t+sl}{1+stl} = \frac{(1-s)(s+t)}{(1+s)(1-st)} = \frac{(1-s)(1-l)}{(1+s)(1+l)}\,.\nnm
\eea
\qed
\rem Note that, in contrast to the 2 dimensional Ising model, this self--dual manifold does not
coincide with the critical manifold \cite{We,Pf}. For example, in the $\Js=\Jt$ plane, the
self--dual line and the critical line coincide only when $\Jst\leq\Js$, then the critical line
splits into 2 dual components. See section \req{Sec_ineq_comp} for an estimate
on the location of these lines.
\section{The Random--cluster model}
\setcounter{equation}{0}
In this section, we introduce the generalized random--cluster model (GRC) and show its connection
to the usual random--cluster model and to the Ashkin--Teller model.

We introduce the model by discussing successively the configuration space, the a
priori measure (generalized percolation measure) and the generalized
random--cluster measure.
\subsection{The model}
\label{sec_GRCmodel}
{\bf Configuration space}

For every bond $b$, let $\Upsilon_b \isdefby \{0,1\}\times\{0,1\}$.\\
The {\defsty configuration space} is the product space $\Upsilon \isdefby 
\Upsilon_b^{\cB(\bbbz^2)}$, where $\cB(\bbbz^2)$ is the set of bonds of
$\bbbz^2$.

A {\defsty configuration of bonds} $\un$ is an element of the configuration
space.
The value of the configuration $\un = (\uns,\unt)$ at a bond $b$ will be denoted either
$n(b) = (\ns(b),\nt(b))$, or $n_b=(\nsb,\ntb)$.

Bonds $b$ such that $n_\s(b)=1$ are said to be $\s$-{\defsty open},
while bonds $b$ such that $n_\s(b)=0$ are said to be $\s$-{\defsty closed}. In the same way we
define {\defsty $\t$-open} and {\defsty $\t$-closed} bonds.

If $\un=(\uns,\unt)$ is some configuration of bonds then ${\overline \un}$ is the configuration given by
${\overline n}_b = (1 - n_{\s,b},1 - n_{\t,b})$.

Let $\un \in \Upsilon$. We define a notion of connectedness for sites, given
the configuration $\un$.\\ The site $i$ is {\defsty $\s$-connected} to the site
$j$, given the configuration $\un$, if there exists a sequence
$t_0=i,t_1,...,t_{k-1},t_k=j$ of sites such that $n_\s(\nnb{t_i}{t_{i+1}})=1,
\forall i=0,...,k-1$.\\
Maximal connected components of sites are called {\defsty $\s$-clusters}. The
number of $\s$-clusters in a configuration $\un$ which intersect a given set
$\Lambda$ is denoted by $N_\s(\un|\Lambda)$; note that each isolated site is a
cluster.\\
Two sets are $\s$-connected, given a configuration $\un$, if there is a point of
the first set which is $\s$-connected to a point of the second set.\\
If $i$ and $j$ are $\s$-connected,we will write
\be
i \sconn j
\ee
We make the corresponding definitions for $\t$.

{\bf The a priori measure}

On the configuration space we introduce an a priori measure, which we call
generalized percolation measure (GP measure).

We introduce for each $b\in\cB(\bbbz^2)$ a probability measure $\mugp_b$ on $\Upsilon_b$,
given by
\bea\label{a_priori}
\mugp_b((0,0)) = a_0(b)\,,&& \mugp_b((1,1)) = a_{\s\t}(b)\,,\nnm\\
\mugp_b((1,0)) = a_\s(b)\,,&& \mugp_b((0,1)) = a_\t(b)\,.
\eea

Let $\cB$ be a finite subset of $\cB(\bbbz^2)$. The {\defsty generalized percolation
measure in $\cB$} is defined as the following product measure on $\Upsilon$
\be
\mugp_\cB(\un) = \prod_{b \in \cB: \atop n_b=(0,0)}a_0(b)
\prod_{b \in \cB: \atop n_b=(1,0)}a_\s(b)\prod_{b \in \cB: \atop n_b=(0,1)}a_\t(b)
\prod_{b \in \cB: \atop n_b=(1,1)}a_{\s\t}(b)\,.
\ee

{\bf The generalized random--cluster measure}

Let $\Lambda$ be a bounded simply connected subset of $\bbbz^2$.\\
We recall that
\be
\cB(\Lambda) \isdefby \{b\in\bbbl:\delta b\subset\Lambda\}\,.
\ee
We introduce another set of edges associated to the set of sites $\Lambda$,
\be
\cB^+(\Lambda) \isdefby  \{b\in \bbbl : \delta b \cap \Lambda \neq \eset\}\,.
\ee
We introduce two kinds of boundary conditions.\\
The configuration $\un$ satisfies the {\defsty $(+,+)$-boundary condition on $\Lambda$} if
\be
n(b) = (1,1)\;\;\forall b \notin \cB^+(\Lambda)\,.
\ee
The configuration $\un$ satisfies the {\defsty free boundary condition on $\Lambda$} if
\be
n(b) = (0,0)\;\;\forall b \notin \cB(\Lambda)\,.
\ee
Notice the fact that the set of bonds in each of these definition is different.
\begin{tabbing}
\rem \=2) \=\kill
\rem 1) $(+,+)$-boundary condition corresponds to what is usually called\\{\it wired} boundary
condition.\\
\>2) We can also define more complicated kinds of boundary conditions\\by imposing the
corresponding values for the configuration outside $\cB(\Lambda)$ or $\cB^+(\Lambda)$.
\end{tabbing}

We introduce the following notations
\bea\label{sumplus}
&\sum_{+,\Lambda} \isdefby \sum_{\un \,:\, \un \mbox{ \scriptsize satisfies the
$(+,+)$-b.c. on } \Lambda}&\nnm\\
&\sum_{f,\Lambda} \isdefby \sum_{\un \,:\, \un \mbox{ \scriptsize satisfies the
free b.c. on } \Lambda}&
\eea

The {\defsty generalized random--cluster measure with $(+,+)$-boundary condition on $\Lambda$}
is the probability measure on $\Upsilon$ given by
\be
\mugrcp(\un|q_\s,q_\t)\isdefby \cases{
\frac{\mugp_{\cB^+(\Lambda)}(\un) q_\s^{\Ns(\un|\Lambda)} q_\t^{\Nt(\un|\Lambda)}}
{\sum_{+,\Lambda}\mugp_{\cB^+(\Lambda)}(\un)q_\s^{\Ns(\un|\Lambda)} q_\t^{\Nt(\un|\Lambda)}}
&  if $\un$ satisfies
the $(+,+)$-b.c. on $\Lambda$,\cr
0& otherwise.\cr}
\ee
where $q_\s$ and $q_\t$ are two positive real numbers.\\
The {\defsty generalized random--cluster measure with free boundary condition on $\Lambda$}
is the probability measure on $\Upsilon$ given by
\be
\mugrcf(\un|q_\s,q_\t)\isdefby \cases{
\frac{\mugp_{\cB(\Lambda)}(\un) 
 q_\s^{\Ns(\un|\Lambda)}q_\t^{\Nt(\un|\Lambda)}}{\sum_{f,\Lambda}\mugp_{\cB(\Lambda)}(q_\s^{\Ns(\un|\Lambda)}
 q_\t^\Nt{(\un|\Lambda)}}
&  if $\un$ satisfies
the free b.c. on $\Lambda$,\cr
0& otherwise.\cr}
\ee

{\bf Relation to the usual percolation and random--cluster measure}

For special classes of functions, which we define below, the expectation value
in the GP or GRC measures can be related to the expectation value in some
percolation or random--cluster measures.

We introduce three classes of functions on $\Upsilon$.\\
Let $\cF^1$ be the set of functions on $\{0,1\}^{\cB(\bbbz^2)}$, and
$\cF^2$ be the set of functions on $[\{0,1\}\times\{0,1\}]^{\cB(\bbbz^2)}$. We define
\bea
&{\cF}_\s \isdefby  \{f\in\cF^2 : \exists f_\s\in \cF^1 \mbox{ with } f(\un) =
f_\s(\uns)\,\,\forall \un\}\,,&\nnm\\
&{\cF}_\t \isdefby  \{f\in\cF^2 : \exists f_\t\in \cF^1 \mbox{ with } f(\un) =
f_\t(\unt)\,\,\forall \un\}\,,&\\
&{\cF}_b \isdefby  \{f\in\cF^2 : \exists f_b\in \cF^1 \mbox{ with } f(\un) =
f_b(\uns \vee \unt)\,\,\forall \un\}\,.&\nnm
\eea
where $(\uns \vee \unt)(b) \isdefby  \max(\ns(b),\nt(b))$.

We prove now an elementary lemma relating the GP measure on $\cB$ to the usual
percolation measure on $\cB$, which is defined on $\{0,1\}^{\cB(\bbbz^2)}$ by
\be
\mup_\cB(\un|\underline{p}) \isdefby  \prod_{b \in \cB: \atop n_b=1}p(b)\prod_{b\in\cB: \atop n_b=0}(1-p(b))\,,
\ee
where $0\leq p(b) \leq 1\;\forall b$.
\begin{lem}\label{lem_comp_perc}
1. If $f \in {\cF}_\s$ then $\mup_\cB(f_\s|p(b)=a_\s(b)+a_{\s\t}(b)) = \mugp_\cB(f)$.\\
2. If $f \in {\cF}_\t$ then $\mup_\cB(f_\t|p(b)=a_\t(b)+a_{\s\t}(b)) = \mugp_\cB(f)$.\\
3. If $f \in {\cF}_b$ then $\mup_\cB(f_b|p(b)=a_\s(b)+a_\t(b)+a_{\s\t}(b)) = \mugp_\cB(f)$.
\end{lem} 
\prf
We have (omitting the dependence on $b$ of the probabilities)
\bea
\mugp_\cB(f) &=& \sum_{\uns}f_\s(\uns)\sum_{\unt}\prod_{b\in\cB} a_0^{\onsb\ontb}
a_\s^{n_{\s,b}\ontb}
a_\t^{\onsb n_{\t,b}}a_{\s\t}^{n_{\s,b} n_{\t,b}}\nnm\\
&=& \sum_{\uns}f_\s(\uns)\prod_{b\in\cB}\sum_{n_{\t,b} = \pm 1}a_0^{\onsb\ontb}a_\s^{n_{\s,b}\ontb}
a_\t^{\onsb n_{\t,b}}a_{\s\t}^{n_{\s,b} n_{\t,b}}\nnm\\
&=& \sum_{\uns}f_\s(\uns)\prod_{b\in\cB} \left(a_0^{\on_{\s,b}}a_\s^{n_{\s,b}}+a_\t^{\on_{\s,b}}
a_{\s\t}^{n_{\s,b}}\right)\nnm\\
&=& \sum_{\uns}f_\s(\uns)\prod_{b\in\cB: \atop \nsb=1} (a_\s+a_{\s\t})\prod_{b\in\cB: \atop \nsb=0}
(a_0+a_\t)\,.
\nnm
\eea
\qed

The two generalized random--cluster measures are also related to the
corresponding usual random--cluster measure on $\Lambda$, which are defined on
$\{0,1\}^{\cB(\bbbz^2)}$ by (using notations similar to \req{sumplus})
\bea
\murc^w(\un|\underline{p},q)\isdefby \cases{
\frac{\mup_{\cB^+(\Lambda)}(\un|\underline{p}) q^{N(\un|\Lambda)}}
{\sum_{+,\Lambda}\mup_{\cB^+(\Lambda)}(\un|\underline{p})q^{N(\un|\Lambda)}}
&  if $\un$ satisfies
the wired b.c. on $\Lambda$,\cr
0& otherwise.\cr}\nnm\\
\murc^f(\un|\underline{p},q)\isdefby \cases{
\frac{\mup_{\cB(\Lambda)}(\un|\underline{p}) q^{N(\un|\Lambda)}}
{\sum_{f,\Lambda}\mup_{\cB(\Lambda)}(\un|\underline{p})q^{N(\un|\Lambda)}}
&  if $\un$ satisfies
the free b.c. on $\Lambda$,\cr
0& otherwise.\cr}
\eea
where $N(\un|\Lambda)$ is the number of clusters in $\un$ intersecting
$\Lambda$.
This is proved in the following
\begin{lem}
\bea
f \in \cF_\s &\Rightarrow& \mugrc^\circ(f|q_\s,1) = \murc^\bullet(f_\s|p=a_\s+a_{\s\t},q_\s)\,,\nnm\\
f \in \cF_\t &\Rightarrow& \mugrc^\circ(f|1,q_\t) = \murc^\bullet(f_\t|p=a_\t+a_{\s\t},q_\t)\,,\nnm
\eea
where $\circ$ means free (resp. wired) boundary condition for the usual random--cluster model, and
$\bullet$ means free (resp $(+,+)$) boundary condition for the GRC model.
\label{lemmacomp}
\end{lem} 
\prf
As $\Ns \in \cF_\s$ we have, by lemma \ref{lem_comp_perc}, (and omitting the dependence on
$b$ of the probabilities)
\bea
\mugrc^\circ(f|q_\s, 1) &=&
\frac{\sum_{\circ,\Lambda}f(\un)\mugp_{\cB^\circ(\Lambda)}(\un) q_\s^{\Ns(\un|\Lambda)}}
{\sum_{\circ,\Lambda}\mugp_{\cB^\circ(\Lambda)}(\un)q_\s^{\Ns(\un|\Lambda)}}\nnm\\
&=& \frac{\sum_{\bullet,\Lambda}f_\s(\un)\mup_{\cB^\bullet(\Lambda)}(\un|p=a_\s+a_{\s\t})
q_\s^{N(\un|\Lambda)}}
{\mup_{\cB^\bullet(\Lambda)}(\un|p=a_\s+a_{\s\t}) q_\s^{N(\un|\Lambda)}}\nnm\\
&=& \murc^\bullet(f_\s|p=a_\s+a_{\s\t},q_\s)\,.\nnm
\eea
\qed
\subsection{Relation to the Ashkin--Teller model}
The Ashkin--Teller model defined in section \ref{sec_ATmodel} and the
generalized random--cluster model defined in the section \ref{sec_GRCmodel} are
closely related as is shown in the following \begin{pro}\label{GRC_AT}
Let
\bea
a_0 &=& \e^{-2(\Js+\Jt)}\,,\nonumber\\
a_\s &=& \e^{-2\Jt}(\e^{-2\Jst}-\e^{-2\Js})\,,\nonumber\\
a_\t &=& \e^{-2\Js}(\e^{-2\Jst}-\e^{-2\Jt})\,,\nonumber\\
a_{\s\t} &=& 1-\e^{-2(\Js+\Jst)}-\e^{-2(\Jt+\Jst)}+\e^{-2(\Js+\Jt)}\,.\label{aaaa}
\eea
The constants $a_0$, $a_\s$, $a_\t$, $a_{\s\t}$ define a probability measure
\req{a_priori} on $\{0,1\}\times\{0,1\}$ if, and only if,
\bea
&\Js \geq \Jst\,,\;\Jt \geq \Jst\,,&\nnm\\
&\Js \geq 0\,,\phm \Jt \geq 0\,,\phm \th\Jst \geq -\th\Js\th\Jt\,.&\nnm
\eea
Moreover, with this choice of probabilities,\\
1.
\bea
&\Zpp = C_2\sum_{+,\Lambda}\mugp_{\cB^+(\Lambda)}(\un) 2^{\Ns(\un|\Lambda)}
2^{\Nt(\un|\Lambda)}\,,\label{fk1}\\
&\Zf =  C_3\sum_{f,\Lambda}\mugp_{\cB(\Lambda)}(\un) 2^{\Ns(\un|\Lambda)}
2^{\Nt(\un|\Lambda)}\,.\label{fk2}
\eea
2.
\bea
&\mupp(\s_A\t_B) = \mugrcp(\kappa^\s_A\kappa^\t_B|2,2)\label{fk3}\,,\\
&\muf(\s_A\t_B) = \mugrcf(\kappa^\s_A\kappa^\t_B|2,2)\label{fk4}\,.
&\eea
where $C_2$, $C_3$ are some constants independent on the configuration and
$\kappa^\s_A$ is the characteristic function which is one on
the configurations such that no finite $\s$-cluster contains an
odd number of sites of $A$; $\kappa^\t_A$ is the corresponding characteristic
function for $\t$. Finally, $\s_A\isdefby\prod_{i\in A}\s_i$ and $\t_B\isdefby\prod_{i\in B}
\t_i$.
\end{pro}
\prf
1. Let us first show that $a_0$, $a_\s$, $a_\t$, $a_{\s\t}$ define a probability measure
on $\{0,1\}\times\{0,1\}$.\\
By definition, $a_0+a_\s+a_\t+a_{\s\t}=1$. Hence we just have to check their positivity\\
Evidently $a_0 \geq 0$.
\bea
&a_\s \geq 0 \Leftrightarrow \Js \geq \Jst\,,&\nnm\\
&a_\t \geq 0 \Leftrightarrow \Jt \geq \Jst\,,&\nnm
\eea
and
\be
a_{\s\t} \geq 0 \Leftrightarrow \e^{-2\Jst}\leq\frac{\ds 1+\e^{-2(\Js+\Jt)}}{\ds\e^{-2\Js}+
\e^{-2\Jt}}
\Leftrightarrow \frac{\ds 1-\e^{-2\Jst}}{\ds 1+\e^{-2\Jst}}\geq -
\frac{\ds 1-\e^{-2\Js}}{\ds 1+\e^{-2\Js}}\frac{\ds 1-\e^{-2\Jt}}{\ds 1+\e^{-2\Jt}}\,,\nnm
\ee
but this is just $\th \Jst \geq -\th\Js\th\Jt$.\\
Now note that
\be
\Js\geq\Jst\,,\;\Jt\geq\Jst\,,\;\th \Jst \geq -\th\Js\th\Jt \Rightarrow \Js\geq 0\,,\;\Jt\geq 0\,.
\ee
Indeed, suppose $\Js<0$ and $\Jt\geq 0$, then $\Jst$ must be positive and therefore larger
than $\Js$ which is a contradiction. If $\Js<0$ and $\Jt <0$ then in this case
\be
\th |\Jst|<\th |\Js| \th |\Jt| < (\th |\Jst|)^2\,,
\ee
which is also a contradiction.

We show that the partition functions can be expressed in term of the denominator of the
generalized random--cluster measures.\\
The weight in the partition function can be expanded in the following way
\bea
\lefteqn{\exp\{\Js \s_i\s_j + \Jt \t_i\t_j + \Jst \s_i\s_j\t_i\t_j\}}\nonumber\\
&=& C \exp\{(\Js+\Jst) (\s_i\s_j-1) + (\Jt+\Jst) (\t_i\t_j-1) + \Jst (\s_i\s_j-1)(\t_i\t_j-1)\}
\nonumber\\
&=& C \{a_0 + a_\s \delta_{\s_i\s_j} + a_\t \delta_{\t_i\t_j} + a_{\s\t} \delta_{\s_i\s_j}
\delta_{\t_i\t_j}\}\,,\label{straight2}
\eea
where $C$ is some constant.
The partition function with free boundary condition can then be written
\bea
\Zf &=& C_3 \sum_{f,\Lambda}\mugp_{\cB(\Lambda)}(\un)
\sum_{\s,\t}\prod_{b=\nnb{i}{j}\in\cB(\Lambda): \atop n_{\s,b}=1}
\delta_{\s_i\s_j}\prod_{b=\nnb{i}{j}\in\cB(\Lambda): \atop
n_{\t,b}=1}\delta_{\t_i\t_j}\nnm\\
&=& C_3 \sum_{f,\Lambda}\mugp_{\cB(\Lambda)}(\un) 2^{\Ns(\un|\Lambda)} 2^{\Nt(\un|\Lambda)}\,.
\eea
The case of $(+,+)$-boundary condition is treated in the same way.

2. We finally prove \req{fk3} and \req{fk4}.\\
The same expansion as above can be done on the correlation functions. We then obtain
\bea
\muf(\s_A\t_B) &=& \frac{\sum_{f,\Lambda}\mugp_{\cB(\Lambda)}(\un) 
\sum_{\s,\t}\s_A\t_B
 \prod_{b=\nnb{i}{j}\in\cB(\Lambda):
\atop n_{\s,b}=1}
\delta_{\s_i\s_j} \prod_{b=\nnb{i}{j}\in\cB(\Lambda): \atop
n_{\t,b}=1}\delta_{\t_i\t_j}}
{\sum_{f,\Lambda}\mugp_{\cB(\Lambda)}(\un) 2^{\Ns(\un|\Lambda)}2^{\Nt(\un|\Lambda)}}\nnm\\
&=& \frac{\sum_{f,\Lambda}\mugp_{\cB(\Lambda)}(\un) \kappa^\s_A(\un)
\kappa^\t_B(\un) 2^{\Ns(\un|\Lambda)}2^{\Nt(\un|\Lambda)}}
{\sum_{f,\Lambda}\mugp_{\cB(\Lambda)}(\un) 2^{\Ns(\un|\Lambda)}2^{\Nt(\un|\Lambda)}}\nnm\\
&=& \mugrcf(\kappa^\s_A\kappa^\t_B|2,2)\,,
\eea
where we have used the fact that the only configurations that will give a non zero
contribution must be such that $\s_A = \t_B = 1$, $\forall \s,\t$. But this is only possible
if the intersection of $A$ and any $\s$-cluster, as well as the intersection of $B$ and any
$\t$-cluster contains an even number of sites.\\
The case of $(+,+)$-boundary condition is treated in the same way, using also the fact that
the sites belonging to the infinite cluster have the fixed value (1,1).

\qed
{\bf Remarks:} 1. As already stated previously, if the conditions on the order of the
coupling constants of the Ashkin--Teller model are not satisfied, then it is still
possible to use the random--cluster representation. Indeed, by first doing a
change of variables, we can always write down the model in the required
form. As an important example, consider the case $\Js=\Jt\leq\Jst$. The change
of variables $(\s,\t) \mapsto (\theta,\t)$ results in
$J_\theta\geq\Jt=J_{\theta\tau}$ to which we can apply the GRC representation.
Notice that the resulting random--cluster measure has the property that
$a_\tau=0$. This implies that the $\theta$-open bonds play the role of the
(random) graph on which the $\tau$-bonds ``live''. We will return to this
particular case later. 

As an important particular case of the above proposition, we have
\bea
\mupp(\s_i) = \mugrcp(i\conn^\s\Lambda^c)\,,\\
\mupp(\s_i\s_j)= \mugrcp(i\conn^\s j)\,.
\eea
\\
2. In fact the generalized random--cluster model with $q_\s$, $q_\t \in \bbbn$ can also be related to
some spin model. More precisely, if we consider the model defined in the following way:
\be\label{HamPotts}
\cH = -2(\Js-\Jst)\delta_{\s_i\s_j}-2(\Jt-\Jst)\delta_{\t_i\t_j}-4\Jst\delta_{\s_i\s_j}
\delta_{\t_i\t_j}\,,
\ee
where $\s_i \in \{1,...,q_\s\}$ and $\t_i \in \{1,...,q_\t\}$, then an analogous proposition as 
the one above still holds. These models are usually called ($q_\s$,
$q_\t$){\em --cubic models} \cite{DR};
they may be thought of as resulting from two coupled Potts models. In the case $\Jt = \Jst$ we recover
the partially symmetric Potts model \cite{DuLaMeMiRu,LaMaRu}. Notice also that the Hamiltonian \req{HamPotts}
cannot be cast into the form of the Potts models considered by Grimmett
\cite{Gr}.\\
3. More complicated boundary condition can be treated in exactly the same way as here.

We are now going to show that the generalized random--cluster model is self-dual.
\subsection{Some geometrical results}
\label{geom_res}
In section \ref{sec_ATmodel} a definition of dual set was introduced. There, the relevant
variables were spins and so the primary geometrical objects were sites. The dual set was
therefore defined starting from those sites, building the corresponding dual pla\-quettes and
completing the cell-complex thus obtained.\\
We are now going to give another notion of dual of a set. As in the random--cluster model
the variables are the bonds, it will be natural to build the dual set starting from
the dual objects associated with the bonds, namely the dual bonds.

Let $\cB$ be a set of bonds. We define an associated cell-complex
$\Lambda(\cB)$: its set of bonds $\Lambda_1(\cB)$ is the set of the
bonds in $\cB$; its set of sites $\Lambda_0(\cB)$ is the set of the boundaries of these bonds;
the set of plaquettes $\Lambda_2(\cB)$ is the set of the plaquettes whose boundary
belongs to $\cB$ (there may be none).

Let $\cB$ be such that $\Lambda_0(\cB)$ is a bounded and simply connected.\\
We define the dual of the set of bonds $\cB$:
\be
{\widehat \cB} = \{\hat{b}\in \bbbl^*: \hat{b}\mbox{ crosses some }b \in \cB\}\,.
\ee
The corresponding dual cell--complex is ${\widehat \Lambda}:=\Lambda(\widehat \cB)$.

Let $\un \in \{0,1\}^{\cB}$ be a configuration of bonds. We define the dual
configuration $\uhn\in \{0,1\}^{\widehat \cB}$ to be
\be
\hn_{\hat{b}} = 1-n_b\,.
\ee
where $\hat{b}$ is the bond of $\widehat \cB$ intersecting $b$.

For a given configuration of bonds $\un$ we denote by $(\Lambda, \un)$ the
graph whose vertex are the sites of $\Lambda(\cB)$ and whose edges are the
open bonds of $\un$.

We then have the following two relations:
\bea
&&N(\un) = |\Lambda|-|\un|+l(\un)\label{Euler}\,,\\
&&N(\un) = l(\uhn)+1\label{TopDual}\,,
\eea
where $N(\un)$, $|\Lambda|$, $|\un|$ and $\l(\un)$ are respectively the
number of connected components, the number of vertices, the number of edges
and the cyclomatic number\footnote{An {\defsty elementary cycle} of an oriented
graph $(V,E)$ (i.e. a graph whose edges have an orientation) is a sequence of
distinct edges $(e_1,\dots,e_n)$ such that every $e_k$ is connected to
$e_{k-1}$ by one of its extremities and to $e_{k+1}$ by the other one (
$e_0 := e_n$, $e_{n+1}:=e_1$) and no vertex of the graph belongs to more
than two of the edges of the family. To each cycle one can associate a vector
$\underline c$ in $\bbbr^{|E|}$ by
$$
c_e:=\cases{\phantom{-}0 & if the edge $e$ does not belong to the cycle,\cr
\phantom{-}1 & if the edge $e$ belongs to the cycle and is positively oriented,\cr
-1 & if the edge $e$ belongs to the cycle and is not positively
oriented.}
$$
A family of elementary cycles is {\defsty independent} if the corresponding
vectors are linearly independent. The {\defsty cyclomatic number} of the
graph is the maximal number of independent elementary cycles of the graph; it
is independent of the orientation.}
of the graph $(\Lambda, \un)$; $\l(\uhn)$ is the cyclomatic number of the
graph $(\widehat\Lambda, \uhn)$.

Relation \req{Euler} is just the well-known Euler formula for the graph
$(\Lambda, \un)$ and can be easily proved (see, for example, Theorem 1
in \cite{Be}). Relation \req{TopDual} becomes clear once we use the fact that
the cyclomatic number of a planar graph $(V, E)$ also corresponds to the
number of bounded connected components of
$\bbbr^2$ delimited by the edges of the graph
(which are called {\em finite faces} in \cite{Be}; see, for example, Theorem 2
therein). Then \req{TopDual} amounts to saying that to each finite cluster
of $\un$ corresponds one and only one such finite component of $\uhn$, which
is straightforward to prove.

Below, we will use an extension of these relations in the case of infinite
graphs $(\bbbz^2,\un)$, where $\un$ is a configuration satisfying the
$+$-boundary conditions. To make sense of the above formul{\ae}, we will
apply them to the restriction of this graph to the graph $(V,B)$, where
$V := \{t\in\bbbz^2\,:\,d_1(t,\Lambda)\leq 1\}$ and $B=B(V)$. This will give
us the relation we require, up to some constant independent of $\un$.

\rem We have only considered simply connected sets $\Lambda$. Let us make
a comment in the case 
of non--simply connected $\Lambda$. To be specific, suppose the set $\Lambda$ 
is a square with some hole in the middle, and that we have $(+,+)$-boundary
conditions in the spin model. Then the associated random--cluster
measure can be defined similarly as was done above, but the corresponding
$+$-boundary condition are such that the two disjoint components of the
the set $\{b\,:\,b\not\in \cB^+(\Lambda)\}$ must be connected by an extra
open bond. This makes the graph {\em non planar} and the relation \req{TopDual}
does not hold anymore. Thus, as was the case for the duality of the spin system,
the condition that $\Lambda$ is simply connected is essential. More general
settings can be studied using techniques of algebraic topology as in
\cite{LaMeRu}.
\subsection{Duality in the random--cluster representation}
\label{DualRC}
Let $\Lambda$ be a bounded simply connected subset of $\bbbz^2$.
 
Let $\un=(\uns,\unt)$ be
a configuration of $\s$- and $\t$-bonds. The dual configuration is defined as
\bea
\hn_{\s,{\hat{b}}} = 1- n_{\t,b}\,,\nonumber\\
\hn_{\t,{\hat{b}}} = 1- n_{\s,b}\,.
\eea
We emphasize the fact that we have exchanged $\s$ and $\t$ bonds, this is done for later
convenience.

Note that if $\un$ satisfies the $(+,+)$- (resp. free) boundary condition on
$\Lambda$, then
$\uhn$ satisfies the free (resp $(+,+)$-) boundary condition on $\Lambda^*$, and that
$\widehat{\cB^+(\Lambda)}=\cB(\Lambda^*)$. Indeed,
let's consider what happens to a single site of $\Lambda$ during the process of going
to the random--cluster representation and then to its dual.

\begin {figure} [htb]
 \centerline{\psfig{figure=\psdir/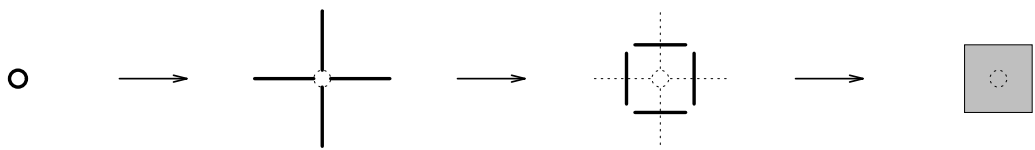,height=18mm}}
 \caption[]{Lattice transformations (only one site shown)}
 \label{fig_duality}
\end{figure}
We start with the Ashkin--Teller model in $\Lambda$ with $(+,+)$-boundary condition. Let us consider
some site $t\in\Lambda$. We then
define the corresponding GRC model with $(+,+)$-boundary condition: here the bonds whose value
is not fixed are all bonds whose boundary contains at least one site of $\Lambda$ (among them, there
are in particular the four bonds with an endpoint at $t$).
Then we take its dual model with free
boundary condition: now, the bonds whose value is not fixed are the dual bonds crossing the
previous ones. 
But the four dual bonds around $t$ are the boundary of a plaquette $p^*$ having $t$ as its center
(see fig \ref{fig_duality}). Hence,
to each $t\in \Lambda$, this associates a plaquette $p^*(t)$ having $t$ as its center: this is just the
definition we have given of $\Lambda^*$. Therefore these four bonds
belong to $\cB(\Lambda^*)$. Doing this for all sites of $\Lambda$,
we obtain all of $\cB(\Lambda^*)$.

Using the preceding geometrical results, we can write
\bea
&&\hspace{-0.5cm}\sum_{+,\Lambda}f(\un)\mugp_{\cB^+(\Lambda)}(\un)
q_\s^{\Ns(\un|\Lambda)} q_\t^{\Nt(\un|\Lambda)} =\nnm\\
&&\;\;=\sum_{\un : \atop  (+,+)\mbox{\tiny -b.c. in }\Lambda}f(\un) \cdot \biggl(\prod_{\sc b\in \cB^+(\Lambda): \atop
\sc n_b=(0,0)}a_0(b)
\prod_{\sc b\in\cB^+(\Lambda): \atop \sc n_b=(1,0)}a_\s(b)\prod_{\sc b\in\cB^+(\Lambda): \atop 
\sc n_b=(0,1)}a_\t(b)
\prod_{\sc b\in\cB^+(\Lambda): \atop \sc n_b=(1,1)}a_{\s\t}(b)\biggr)\times\nnm\\
&&\;\;\;\;\;\times q_\s^{\Ns(\un|\Lambda)}q_\t^{\Nt(\un|\Lambda)}\nnm\\
&&\;\;=C\sum_{\uhn: \atop \mbox{\tiny free b.c. in }\Lambda^*}f(\un) \cdot \biggl(\prod_{\sc \hb\in \cB(\Lambda^*):
\atop \sc \hn_\hb=(1,1)}a_0(b)
\prod_{\sc\hb\in \cB(\Lambda^*): \atop \sc\hn_\hb=(1,0)}a_\s(b)
\prod_{\sc\hb\in \cB(\Lambda^*): \atop \sc\hn_\hb=(0,1)}a_\t(b)
\prod_{\sc\hb\in \cB(\Lambda^*): \atop \sc\hn_\hb=(0,0)}a_{\s\t}(b)\biggr)\times\nnm\\
&&\;\;\;\;\;\times q_\s^{\Nt(\uhn|\Lambda)+|\uhnt|}q_\t^{\Ns(\uhn|\Lambda)+|\uhns|}\nnm\\
&&\;\;=C\sum_{\uhn: \atop \mbox{\tiny free b.c. in }\Lambda^*}f(\un) \cdot \biggl(\prod_{\sc\hb\in \cB(\Lambda^*):
\atop \sc\hn_\hb=(1,1)}\ha_{\s\t}(\hb)
\prod_{\sc\hb\in \cB(\Lambda^*): \atop \sc\hn_\hb=(1,0)}\ha_\s(\hb)
\prod_{\sc\hb\in \cB(\Lambda^*): \atop \sc\hn_\hb=(0,1)}\ha_\t(\hb)
\prod_{\sc\hb\in \cB(\Lambda^*): \atop \sc\hn_\hb=(0,0)}\ha_0(\hb)\biggr)\times\nnm\\
&&\;\;\;\;\;\times \widehat{q}_\s^{\Ns(\uhn|\Lambda)}\widehat{q}_\t^{\Nt(\uhn|\Lambda)}\,,
\label{GRCdual}
\eea
where $C$ denote some constant independent on $\un$, $|\uhn|$ is the number of open bonds in $\uhn$
and we have introduced
\bea\label{dualRC}
\ha_0(\hb) &:=& C' a_{\s\t}(b)\,,\nnm\\
\ha_\s(\hb) &:=& q_\t C' a_\s(b)\,,\nnm\\
\ha_\t(\hb) &:=& q_\s C' a_\t(b)\,,\nnm\\
\ha_{\s\t}(\hb) &:=& q_\s q_\t C' a_0(b)\,,\nnm\\
\widehat{q}_\s &:=& q_\t\,,\nnm\\
\widehat{q}_\t &:=& q_\s\,.
\eea
$C'$ is a normalization constant such that 
$\ha_0+\ha_\s+\ha_\t+\ha_{\s\t}=1$:
\be
C'^{-1} = a_{\s\t} + q_\t a_\s + q_\s a_\t + q_\s q_\t a_0\,.
\ee
\begin{tabbing}
\rem \=2) \=\kill
\rem 1) Note that the positivity of theses dual probabilities is a consequence of the\\
positivity of the initial probabilities.\\
\>2) There are no other way to distribute the factors $q_\s$ and $q_\t$ in \req{GRCdual}.
\end{tabbing}

As a consequence we have
\begin{pro}
Let $\Lambda$ be a finite, simply connected subset of $\bbbz^2$. Let $\ha_0$, $\ha_\s$, $\ha_\t$ and
$\ha_{\s\t}$ be defined by \req{dualRC}. Then, for all $f\in\cF^2$,
\be
\mugrcp(f|a_0,a_\s,a_\t,a_{\s\t},q_\s,q_\t) = \mugrcb^f_{\Lambda^*}
(\widehat{f}|\ha_0,\ha_\s,\ha_\t,\ha_{\s\t},\hat{q}_\s,\hat{q}_\t)\,,
\ee
where $\widehat{f}(\uhn) \isdefby f(\un(\uhn))$.
\end{pro}

A natural question to ask is: do the transformations between the spin and random--cluster models
and the duality transformations commute in the Ashkin--Teller case?
The answer is yes, as is shown below.
\subsection{Commutativity of the dualities and RC transformation}
We want to compare the dual of the generalized random--cluster model associated to
the Ashkin--Teller model with $(+,+)$-boundary conditions with the random--cluster
model associated to the dual of this Ashkin--Teller model.

For the dual of the random--cluster model, we obtain (see \req{dualRC} and \req{aaaa})
\bea
\ha_0 &=& \frac{1-j_{\s\t}(j_\s+j_\t)+j_\s j_\t}{1+j_{\s\t}(j_\s+j_\t)+j_\s
j_\t}\,,\nnm\\
\ha_\s &=& \frac{2j_\t(j_{\s\t}-j_{\s})}{1+j_{\s\t}(j_\s+j_\t)+j_\s j_\t}\,,\nnm\\
\ha_\t &=& \frac{2j_\s(j_{\s\t}-j_{\t})}{1+j_{\s\t}(j_\s+j_\t)+j_\s j_\t}\,,\nnm\\
\ha_{\s\t} &=& 1-\ha_0-\ha_\s-\ha_\t\,,
\label{drc}
\eea
where $j_\s = \e^{-2J_\s}$, $j_\t = \e^{-2J_\t}$, $j_{\s\t} = \e^{-2J_{\s\t}}$.

For the random--cluster of the dual model, we find (see \req{DualRel} and \req{aaaa})
\bea
a_0^* &=& \frac{l+st}{1+stl}\,,\nnm\\
a_\s^* &=& \frac{(t-l)(1-s)}{1+stl}\,,\nnm\\
a_\t^* &=& \frac{(s-l)(1-t)}{1+stl}\,,\nnm\\
a_{\s\t}^* &=& 1-a_0^*-a_\s^*-a_\t^*\,.
\label{rcd}
\eea
where $l$, $s$, $t$ have been defined in \req{stl}.
Using the relations:
$$
s = \frac{1-j_\s}{1+j_\s},\phm t=\frac{1-j_\t}{1+j_\t},\phm
l=\frac{1-j_{\s\t}}{1+j_{\s\t}}\,,
$$
it is easy to see that the quantities defined in \req{drc} and \req{rcd} are in fact the same.

We have already checked that both resulting models are defined on the same lattice (see section
\ref{DualRC}), so we can conclude
that the following diagram is commutative:

\begin{center}
\setlength{\unitlength}{0.15\textwidth}%
\begin{picture}(1.5,1.2)
\thinlines
\put(0.2,1){\vector( 1, 0){ 1.1}}
\put(-0.13,0.8){\vector( 0, -1){ 0.6}}
\put(0.2,0){\vector( 1, 0){ 1.1}}
\put(1.65,0.8){\vector( 0, -1){ 0.6}}
\put(0,1){\makebox(0,0)[r]{$AT$}}
\put(1.5,1){\makebox(0,0)[l]{$AT^*$}}
\put(0,0){\makebox(0,0)[r]{$RC$}}
\put(1.5,0){\makebox(0,0)[l]{$RC^*$}}
\put(0.75,1.1){\makebox(0,0){$*$}}
\put(-0.33,0.5){\makebox(0,0){$\cF\cK$}}
\put(1.853,0.5){\makebox(0,0){$\cF\cK$}}
\put(0.75,0.12){\makebox(0,0){$*$}}
\end{picture}
\end{center}

Here $\cF\cK$ denotes the random--cluster transformation and * the dualities.\\

We now turn to the properties of the generalized random--cluster model.
\section{Properties of the GRC measure}
\setcounter{equation}{0}
\subsection{FKG inequalities}
In this section we are going to show that the measures of the generalized random--cluster
model which have been introduced are FKG. No hypothesis on $\Lambda$, except its
boundedness, is required.\\

We partition the bonds into two classes:
\bea
{\cB}_> = \{b : a_{\s\t}(b)a_0 (b)\geq a_\s(b) a_\t(b)\}\,,\nnm\\
{\cB}_< = \{b : a_{\s\t}(b)a_0 (b)< a_\s(b) a_\t(b)\}\,.
\eea
We introduce the following partial order on $\{0,1\}\times\{0,1\}$:
\be\label{order1}
(0,0) \preceq (0,1) \preceq (1,1), \phm\phm(0,0) \preceq (1,0) \preceq (1,1)\,,
\ee
for bonds in $\cB_>$, and
\be
(0,1) \preceq (1,1) \preceq (1,0), \phm\phm(0,1) \preceq (0,0) \preceq (1,0)\,,
\ee
for bonds in $\cB_<$.

For the generalized random--cluster associated to the Ashkin--Teller model, it is easy to see
that all bonds will be in $\cB_>$ if $\Jst \geq 0$ and in $\cB_<$ if $\Jst < 0$.
\begin{defn}
Let $\underline m$ and $\underline n$ be two configurations. $\underline m$ is said to {\rm
dominate} $\underline n$, $\underline m \succeq \underline n$, if $m_b \succeq n_b$, $\forall b$.
\end{defn}
\begin{defn}
A function f is said to be {\rm increasing} if $\underline m \succeq \underline n 
\Rightarrow f(\underline m) \geq f(\underline n)$. It is said to be {\rm decreasing} if 
$-f$ is increasing.
\end{defn}
\ex $\,\Ns(\un|\Lambda)$ and $\Nt(\un|\Lambda)+\sum_{b \in \cB_<}n_{\t,b}$ are decreasing functions for
the order defined above, while \/$\Ns(\un|\Lambda)+\sum_b n_{\s,b}$ and $\Nt(\un|\Lambda)+
\sum_{b \in \cB_>}n_{\t,b}$
are increasing functions.\\
Let us just look at the case $\Nt(\un|\Lambda)+\sum_{b \in \cB_>}n_{\t,b}$. It is sufficient to
consider two configurations $\un \succeq \uhn$, differing only by one $\t$-bond at $b$.\\
There are two cases: either $b \in \cB_>$, or $b \in \cB_<$. In the first case the
$\t$-bond is missing in $\uhn$ and hence we have
$$
\Nt(\uhn|\Lambda)+\sum_{b \in \cB_>}\hn_{\t,b} = \Nt(\un|\Lambda)+\sum_{b \in \cB_>}n_{\t,b} + 
(\Nt(\uhn|\Lambda)-\Nt(\un|\Lambda))-1 \leq \Nt(\un|\Lambda)+\sum_{b \in \cB_>}n_{\t,b}\,,
$$ 
since $|\Nt(\uhn|\Lambda)-\Nt(\un|\Lambda)|\leq 1$.\\
If $b\in\cB_<$, then the bond is missing in $\un$ and
$$
\Nt(\uhn|\Lambda)+\sum_{b \in \cB_>}\hn_{\t,b} = \Nt(\un|\Lambda)+\sum_{b \in \cB_>}n_{\t,b} + 
(\Nt(\uhn|\Lambda)-\Nt(\un|\Lambda)) \leq \Nt(\un|\Lambda)+\sum_{b \in \cB_>}n_{\t,b}\,.
$$
Indeed, the $\t$-bond links two sites already in the same $\t$-cluster and therefore
the number of such clusters doesn't change, or it links two different clusters and 
$\Nt(\uhn|\Lambda)-\Nt(\un|\Lambda)=-1$.\\
\begin{defn}
A measure $\mu$ is said to be {\rm FKG} if $\mu(fg)\geq\mu(f)\mu(g)$, for all increasing
functions $f$ and $g$.
\end{defn}
\begin{lem}
The generalized percolation measure $\mugp_\cB$ is FKG for the partial order introduced above.
\end{lem}
\prf
It is sufficient to check that (see \cite{FoKaGi})
$$
\mugp_\cB(\un \vee \un')\mugp_\cB(\un \wedge \un') \geq \mugp_\cB(\un)
\mugp_\cB(\un')\,,
$$
where $a\vee b$ denotes the least upper bound of $a$ and $b$, while $a\wedge b$ denotes their
greatest lower bound.
As $\mugp_\cB$ is a product-measure, it is sufficient to verify this for each bond,
which is straightforward. The only non-trivial inequalities are:\\
$$\mugp_b((1,1))\mugp_b((0,0)) \geq \mugp_b((1,0)) \mugp_b((0,1))\,,$$
for bonds in $\cB_>$, but this is satisfied by definition of this class of bonds; and
$$\mugp_b((1,0))\mugp_b((0,1)) \geq \mugp_b((1,1)) \mugp_b((0,0))\,,$$
for bonds in $\cB_<$, which is also true by definition.
\qed
\begin{pro}
Suppose $q_\s \geq 1$ and $q_\t \geq 1$. Then the random--cluster measure
is FKG for the partial order introduced above.
\end{pro}
\prf
It is sufficient to check (see \cite{FoKaGi}) that
$$
q_\s^{\Ns(\un \vee \un'|\Lambda)+\Ns(\un \wedge \un'|\Lambda)}q_\t^{\Nt(\un \vee \un'|\Lambda)+
\Nt(\un \wedge \un'|\Lambda)} \geq
q_\s^{\Ns(\un|\Lambda)+\Ns(\un'|\Lambda)}q_\t^{\Nt(\un|\Lambda)+\Nt(\un'|\Lambda)}\,,
$$
which can be proved exactly as in the case of the usual random--cluster model
\cite{AiChChNe}.
The only thing to observe is that
\bea
\Ns(\un {\vee} \un'|\Lambda) &=& \Ns(\un \widehat{\vee} \un'|\Lambda)\,,\nnm\\
\Ns(\un {\wedge} \un'|\Lambda) &=& \Ns(\un \widehat{\wedge} \un'|\Lambda)\,,\nnm\\
\Nt(\un {\wedge} \un'|\Lambda) &=& \Nt(\un \widehat{\vee} \un'|\Lambda)\,,\nnm\\
\Nt(\un {\vee} \un'|\Lambda) &=& \Nt(\un \widehat{\wedge} \un'|\Lambda)\,,\nnm\\
\eea
where $\widehat{\vee}$ and $\widehat{\wedge}$ denote the order induced by setting the order
\req{order1} at all bonds. Hence,
\bea
\Ns(\un {\vee} \un'|\Lambda)+\Ns(\un {\wedge} \un'|\Lambda) &=& \Ns(\un \widehat {\vee} \un'|\Lambda)+
\Ns(\un \widehat{\wedge} \un'|\Lambda)\,,\nnm\\
\Nt(\un {\vee} \un'|\Lambda)+\Nt(\un  {\wedge} \un'|\Lambda) &=& \Nt(\un \widehat{\vee} \un'|\Lambda)+
\Nt(\un \widehat{\wedge} \un'|\Lambda)\,.\nnm
\eea
\qed

As direct elementary applications of these inequalities to the Ashkin--Teller model, we have
\bea
\muAT{\circ}{\s_i\s_j} = \mugrc^{\bullet}(i \conn^\s j) &\geq&  \mugrc^{\bullet}
(i \conn^\s \Lambda^c \mbox{ and } j \conn^\s \Lambda^c)\nnm\\
&\geq& \mugrc^{\bullet}
(i \conn^\s \Lambda^c)\mugrc^{\bullet}(j \conn^\s \Lambda^c) = \muAT{\circ}{\s_i}
\muAT{\circ}{\s_j}\,,
\eea
which is nothing more than one of Griffith's inequalities. Here $\circ$ means
any boundary conditions for the Ashkin--Teller model and $\bullet$ the
corresponding boundary conditions for the random--cluster model.

More interesting is the following inequality,
which holds in the case $J_{\s\t} \leq 0$ (for which we cannot use Griffith's
inequalities),
\be
\mupp(\s_i\s_j\t_k\t_l) = \mugrcp(i \conn^\s j, k \conn^\t l) \leq 
\mugrcp(i \conn^\s j)\mugrcp(k \conn^\t l) = \mupp(\s_i\s_j)
\mupp(\t_k\t_l)\,,\nnm
\ee
where we have used the fact that $\ds k \conn^\t l$ is {\it decreasing} for the order $\preceq$.\\
More generally, we have, for negative $\Jst$,
\bea
\muAT{}{\s_A\s_B} &\geq& \muAT{}{\s_A}\muAT{}{\s_B}\,,\nnm\\
\muAT{}{\t_A\t_B} &\geq& \muAT{}{\t_A}\muAT{}{\t_B}\,,\nnm\\
\muAT{}{\s_A\t_B} &\leq& \muAT{}{\s_A}\muAT{}{\t_B}\,,\nnm
\eea
as can be easily verified (this is true for free, as well as for $(+,+)$-boundary conditions).
\subsection{Comparison inequalities}
\label{Sec_ineq_comp}
There is a class of inequalities in the usual random--cluster model which is
very interesting: they allow one to compare the probability of an event for
different values of $q$ and of the probability of occupation. It is possible to
generalize these inequalities here, as shown now.

We consider two random--cluster measures, $\mugrc$ and $\hmugrc$, with parameters
$a_0$, $a_\s$, $a_\t$, $a_{\s\t}$,  $q_\s$, $q_\t$ and $\ha_0$, $\ha_\s$, $\ha_\t$, $\ha_{\s\t}$, 
$\hq_\s$, $\hq_\t$, respectively. What is the relation between the probabilities of monotonous
events computed with these two measures? We will consider only two cases, but others can
be proved in the same way.

We introduce the following notations
\bea
&\ds \rho_\s = \frac{q_\s}{\hat{q}_\s},\; \ds \rho_\t = \frac{q_\t}{\hat{q}_\t},&\\
&\ds\a0 = \frac{a_0}{\ha_0},\; \ds\as = \frac{a_\s}{\ha_\s},\; \ds\at = \frac{a_\t}{\ha_\t},\;
\ds\ast = \frac{a_{\s\t}}{\ha_{\s\t}}.&
\eea

We can now formulate our first inequality.
\begin{lem}
\label{lemcomp1}
Suppose $\hat{q}_\s$, $\hat{q}_\t \geq 1$, and
\bea
& &q_\s \leq \hat{q}_\s\,,\; q_\t \leq \hat{q}_\t\,,\label{qhq}\\
& &\ds\ast \geq\max(\as,\at) \geq \min(\as,\at) \geq \a0, \forall b \in
\cB_>\,,\\
& &\ds\rho_\t\as \geq\max(\ast,\rho_\t\a0)\geq\min(\ast,\rho_\t\a0)
\geq\at\,,\; \forall b \in \cB_<\,,
\eea
then, for every increasing function $A$,
$$
\mugrc(A) \geq \hmugrc(A)\,.
$$
\end{lem}
\prf
Let $\chi(\un) = \frac{\mugrc(\un)}{\hmugrc(\un)}$. We are going to write $\chi$ in such
a way as to make explicit the monotonicity of this function under the above hypotheses.
\bea
\chi &=& C\left\{\prod_{b \in \cB_>}\left(\frac{a_0}{\ha_0}\right)^{{\overline n}_{\s,b}
{\overline n}_{\t,b}}
\left(\frac{a_\s}{\ha_\s}\right)^{n_{\s,b}{\overline n}_{\t,b}}
\left(\frac{a_\t}{\ha_\t}\right)^{{\overline n}_{\s,b}n_{\t,b}}
\left(\frac{a_{\s\t}}{\ha_{\s\t}}\right)^{n_{\s,b}n_{\t,b}}
\right\}\times\nnm\\
& &\times\left\{\prod_{b \in \cB_<}\left(\frac{q_\t}{\hq_\t}\frac{a_0}{\ha_0}
\right)^{{\overline n}_{\s,b}
{\overline n}_{\t,b}}
\left(\frac{q_\t}{\hq_\t}\frac{a_\s}{\ha_\s}\right)^{n_{\s,b}{\overline n}_{\t,b}}
\left(\frac{a_\t}{\ha_\t}\right)^{{\overline n}_{\s,b}n_{\t,b}}
\left(\frac{a_{\s\t}}{\ha_{\s\t}}\right)^{n_{\s,b}n_{\t,b}}
\right\}\times\nnm\\
& &\times\left(\frac{q_\s}{\hq_\s}\right)^{\Ns(\un|\Lambda)}\left(\frac{q_\t}{\hq_\t}
\right)^{\Nt(\un|\Lambda)+
\sum_{b \in \cB_<} n_{\t,b}}\,,\nnm
\eea
where $C>0$ is a constant independent on the configuration.\\
Now, using \req{qhq} and the fact that $\Ns(\un|\Lambda)$ and $\Nt(\un|\Lambda) + \sum_{b \in
\cB_<}n_{\t,b}$ are decreasing, it is easy to see that $\chi$ will be
increasing if what is inside the brackets is increasing; and this will be true
if it is true for each bond. This can be easily checked. We just consider two
examples, since the other cases can be treated in the same way.\\ Let us first
verify that the expression in the first brackets is not decreasing when $n_b$
increases from $(1,0)$ to $(1,1)$. In the first case this expression equals
${a_\s}/{\ha_\s}$, while in the second it equals
$a_{\s\t}/{\ha_{\s\t}}$, which is not smaller by hypothesis.\\ Let's now
show that the expression in the second brackets does not decrease when $n_b$
increases from $(1,1)$ to $(1,0)$. But this amounts to
$a_{\s\t}/{\ha_{\s\t}} \leq {q_\t a_\s}/{\hq_\t\ha_\s}$ which
is true by hypothesis.\\
Doing the same computation for the other cases, we finally obtain
$$
\mugrc(A)=\hmugrc(A|\chi)\geq\hmugrc(A)\,,
$$
by FKG and the fact that $\hmugrc$ and $\mugrc$ are normalized.
 
\qed
We now give a second inequality,
\begin{lem} 
\label{lemcomp2}
Suppose $\hat{q}_\s$, $\hat{q}_\t \geq 1$, and
\bea
& &q_\s \geq \hq_\s\,,\; q_\t \geq \hq_\t\,,\\
& &\ds\ast \geq \max(\rho_\t\as,\rho_\s\at) \geq \min(\rho_\t\as,\rho_\s\at) \geq \rho_\s
\rho_\t\a0\,,\; \forall b \in \cB_>\,,\\ 
& &\ds\as \geq \max(\ast,\rho_\s\a0) \geq \min(\ast,\rho_\s\a0) \geq\rho_\s\at, \forall b \in
\cB_<\,,
\eea
then, for every increasing function $A$,
$$
\mugrc(A) \geq \hmugrc(A)\,.
$$
\end{lem}
\prf
As above, using the fact that $\Ns(\un|\Lambda)+ |\uns|$ and $\Nt(\un|\Lambda) + \sum_{b \in \cB_>}n_{\t,b}$
are increasing.
\qed

\rem As said before, other such inequalities can be proved in exactly the same way, for example
when $q_\s$ increases but $q_\t$ decreases.

As a simple application of these inequalities, we prove inequalities relating the
generalized random--cluster model to the usual one.
\begin{lem}
Suppose $q_\s$, $q_\t \geq 1$, $f \in \cF_\s$, increasing, then
$$
\murc(f_\s|p_1,q_\s) \leq
\mugrc(f|q_\s,q_\t) \leq
\murc(f_\s|p_2,q_\s)\,,
$$
where 
\bea
& &p_1 = \frac{q_\t a_\s+a_{\s\t}}{q_\t(a_0+a_\s)+a_\t+a_{\s\t}}\,,\nnm\\
& &p_2 = a_\s+a_{\s\t}\,.\nnm
\eea
The same kind of relations holds for $f \in \cF_\t$.
\end{lem}
\prf
$$
\mugrc(f|q_\s,q_\t) \leq \mugrc(f|q_\s,1) = \murc(f_\s|p=a_\s+a_{\s\t},q_\s)\,,
$$
where we used lemma \ref{lemcomp1} and lemma \ref{lemmacomp}.
In a similar way,
\bea
\mugrc(f|q_\s,q_\t) &\geq& \mugrc(f|q_\s,q_\t=1,\ha_0=\frac{q_\t a_0}{N},
\ha_\s=\frac{q_\t a_\s}{N}, \ha_\t=
\frac{a_\t}{N}, \ha_{\s\t}=\frac{a_{\s\t}}{N})\nnm\\
&=& \murc(f_\s|(q_\t a_\s+a_{\s\t})/(q_\t(a_0+a_\s)+a_\t+a_{\s\t}),q_\s)\,,\nnm
\eea
where $N = q_\t(a_0+a_\s)+a_\t+a_{\s\t}$ is the normalization of the new probabilities, and
we used lemma \ref{lemcomp2} and lemma \ref{lemmacomp}.
\qed

\begin {figure} [htb]
 \centerline{\psfig{figure=\psdir/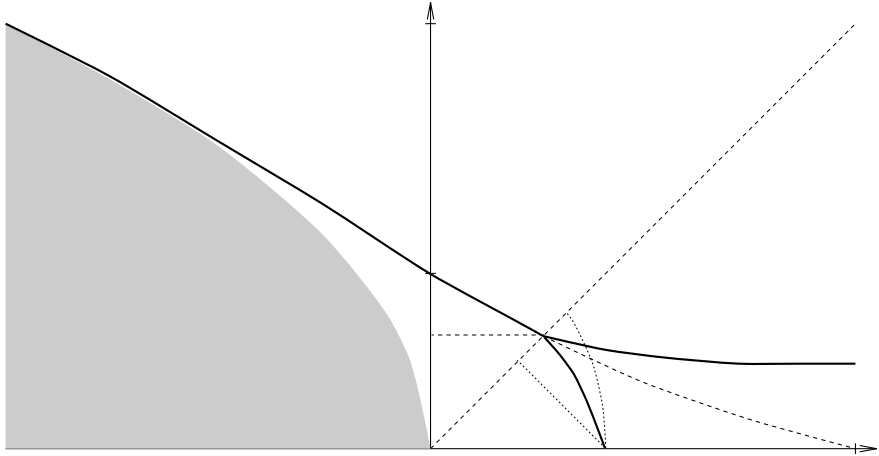,height=60mm}}
 \caption[]{Schematic representation of the phase diagram in the plane $\tanh\Jst$
versus $\tanh\Js=\tanh\Jt$. We have indicated the critical lines (solid lines),
and the self--dual line (which coincides with the solid line up to the splitting
and then follows the dashed line). The shaded region corresponds to the set of
parameters at which the random--cluster representation is not available. The
estimates of the location of the critical line are also shown (dotted lines).}
 \label{fig_diagphase}
\end{figure}

From this lemma we can for example obtain estimates on the location of the critical lines
of the Ashkin--Teller model in the sector $\Jst\geq\Js=\Jt\defines J\geq 0$, in which the critical line
splits into two parts, the first one corresponding to the ordering of the $\s_i$ and $\t_i$, and the
other one to the ordering of the $\theta_i=\s_i\t_i$ (see fig. \ref{fig_diagphase}). Let's study
this second line.

We first make the
change of variables $(\s,\t) \mapsto (\theta,\t)$. The previous lemma then gives, specializing
to the increasing event $\{i \conn^\theta \Lambda^c\}$,
$$
\murc
(i \conn^\theta \Lambda^c|(2 a_\theta+a_{\theta\t})/(1+a_0+a_\theta),2) \leq\mugrc(i \conn^\theta
\Lambda^c|2,2) \leq
\murc(i \conn^\theta \Lambda^c|a_\theta+a_{\theta\t},2)\,,
$$
which can be rewritten in terms of the original Ashkin--Teller and Ising measures:
\be\label{ATcompIsing}
\muIs(\theta_i|J_1) \leq \muAT{}{\theta_i} \leq \muIs(\theta_i|J_2)\,,
\ee
where $\muIs(\cdot|J)$ denotes the Ising  measure with coupling constant $J$, and the coupling
constants $J_1$ and $J_2$ are given by
\bea
J_1 &=& \Jst +\frac{1}{2}\ln\ch(2J)\,,\nnm\\
J_2 &=& J+\Jst\,.\nnm
\eea
The upper bound in \req{ATcompIsing} is easily obtained using GKS inequalities.
This seems not to be the case for the lower bound.
 
Since we know the critical temperature of these two Ising models, we obtain the following
results (remember $\theta_i=\s_i\t_i$):
\bea
&\Jst \leq J^c_{\math{Ising}} -J \Rightarrow  \muATth{}{\s\t}=0\,,&\\
&\Jst > J^c_{\math{Ising}} -\frac{1}{2}\ln\ch(2J) \Rightarrow
\muATth{}{\s\t}>0\,,&
\eea
where $J^c_{\math{Ising}} = \frac{1}{2}\math{\ash} 1$ is Ising critical temperature.\\

\rem Such a behaviour cannot occur in the sector $\Js=\Jt\geq \Jst\geq 0$. Indeed, in this case,
\bea
\muAT{}{\s_i}&=&\mugrc(i \conn^\s \Lambda^c)\geq \mugrc(i \conn^\s \Lambda^c \mbox{  and  }
{i \conn^\t \Lambda^c})= \muAT{}{\theta_i}\nnm\\
 &\geq& \muAT{}{\s_i}\muAT{}{\t_i}=(\muAT{}{\s_i})^2\,,\nnm
\eea
which implies $\muATth{}{\theta_i} =0 \Leftrightarrow \muATth{}{\s_i}=0(=\muATth{}{\t_i})$.
\section{Possible extensions of the model}
\setcounter{equation}{0}
The model above can be extended in several directions.\\
The existence of the random--cluster representation and its properties (except duality) do not
depend on the particular structure of $\bbbz^2$. In fact all this can be shown to remain valid for
an arbitrary finite subgraph $\Lambda$ of some simple graph \cG, applying exactly the same
techniques as those used in this paper.\\
The second direction in which the model may be extended is the following. Suppose
we have $N$ possibly different Potts models on $\Lambda$, interacting through the following
Hamiltonian
\be
\cH = -\sum_{\nnb{i}{j}}\{\sum_{k=1}^N \sum_{r_1<...<r_k} J_k^{(r_1,...,r_k)}
\prod_{t=1}^k(\delta_{\s^{r_t}_i\s^{r_t}_j}-1)\}\,,
\ee
which is an obvious generalization of \req{HamPotts} ($\s_i^k \in \{1,...,q_k\}$).
We can then define a generalized percolation measure ($\un \isdefby (\un_1,...,\un_k)$)
\be
\mugp_\cB(\un) \isdefby  \prod_{A\subset \{1,...,N\}}\prod_{b\in\cB: \atop {n_k(b)=1, \forall
k \in A \atop n_k(b)=0, \forall k \notin A}}a^A\,,
\ee
where $\un_i \in \{0,1\}^{\cB(\Lambda)}$ and
\be
a^A = \exp\bigl(\sum_{k=1}^N\sum_{r_1<...<r_k: \atop r_i \notin A, \forall
i}(-1)^k J_k^{(r_1,...,r_k)}\bigr) -\sum_{B\subset A \atop B \neq A}a^B\,,
\ee
for all $A \subset \{1,...,N\}$.\\
These coefficients will then be positive under suitable conditions on the
coupling constants, so that they can be interpreted as probabilities.\\
It is then possible to define a generalized random--cluster model:
\be
\mugrcb^\star(\un|q_1,\dots,q_N)\isdefby \cases{
\frac{\mugp_{\cB^\star(\Lambda)}(\un) \prod_{k=1}^N q_k^{N_k(\un)}}
{\sum_{\star,\Lambda}\mugp_{\cB^\star(\Lambda)}(\un)\prod_{k=1}^N q_k^{N_k(\un|\Lambda)}}
&  if $\un$ satisfies
the $\star$-b.c. on $\Lambda$,\cr
0& otherwise.\cr}
\ee
where $\star$ denotes boundary condition, and $N_k(\un|\Lambda)$ is the number of clusters of
type $k$ in the configuration $\un$, i.e. $N_k(\un|\Lambda)\isdefby N(\un_k|\Lambda)$.

It will then be possible, introducing enough classes of bonds, to prove again
FKG inequalities and then comparison inequalities.

Again a proposition analogous to proposition \ref{GRC_AT} holds for these new models.
\section{Conclusion}
\setcounter{equation}{0}
In this paper we have defined a generalized random--cluster model and shown how it is related to
the usual random--cluster model and to the Ashkin--Teller model.
This new model still possesses the main properties of the usual random--cluster model,
namely FKG inequalities, comparison inequalities and a duality transformation commuting with the
duality transformation of
the Ashkin--Teller model.\\
Only direct applications of the obtained inequalities have been given (correlation
inequalities, inequalities relating the generalized random--cluster model to the usual one, and
estimates for the critical lines of the Ashkin--Teller model), however many
known results about the random--cluster model can be extended in a
straightforward way. One of our motivations was to develop tools which have
been shown to be very useful in the study of large deviations in the
Ising model (see e.g. \cite{Io,Pi}). 
\section{Appendix}
\setcounter{equation}{0}
{\bf Proof of Proposition \ref{Prop_dual}.}
The equality of the two partition functions follows from relations \req{DualRel} and
comparison of \req{Zpp} and \req{Zf}. Note that the summation is over all families
of (compatible) closed contours without further constraints. This is the case because
$\Lambda$ is simply connected (see section \ref{lowT}).\\
{\bf 1.} We first show that \req{DualRel} is well defined, that is, that the functions $S$, $T$ and
$L$ are strictly positive for any given triple $(\Js^*,\Jt^*,\Jst^*) \in \cD$.\\
This is obvious if $\Jst^*\geq 0$, so that we only consider the case $\Jst^*<0$. In this case,
we have
\bea
&s> st^2>-lt \Leftrightarrow S>0\,,&\nnm\\
&t> ts^2 >-sl \Leftrightarrow T>0\,,&\nnm\\
&l>-st \Leftrightarrow L>0\,.&\nnm
\eea
{\bf 2.} For every triple $(\Js^*,\Jt^*,\Jst^*) \in \cD$, we can solve \req{DualRel} and get
a unique triple $(\Js,\Jt,\Jst)$.\\
{\bf 3.} We now show that the map just defined on $\cD$
\be\label{DualityCC}
(\Js^*,\Jt^*,\Jst^*) \mapsto (\Js,\Jt,\Jst)\,.
\ee
takes its values in $\cD$.\\
{\bf 3.a} $\Js\geq\Jt$\\
$$
\Js\geq\Jt \Leftrightarrow \e^{-2(\Js-\Jt)}\leq 1 \Leftrightarrow T\leq S \Leftrightarrow s(1-l)
\geq t(1-l)\,.
$$
{\bf 3.b} $\Jt>0$\\
$$
\Jt>0 \Leftrightarrow SL < T \Leftrightarrow (1-l^2)ts^2< (1-l^2)t\,.
$$
{\bf 3.c} $\Jt\geq\Jst$\\
$$
\Jt\geq\Jst \Leftrightarrow L \leq T \Leftrightarrow t(1-s)\leq l(1-s)\,.
$$
{\bf 3.d} $\th\Jst>-\th\Js\th\Jt$\\
We use the following elementary result
\be\label{rel_th}
\th a \geq -\th b\;\th c \Leftrightarrow \frac{1-\alpha}{1+\alpha}\geq -
\frac{1-\beta}{1+\beta}\frac{1-\gamma}{1+\gamma} \Leftrightarrow
\alpha(\beta+\gamma) \leq 1+\beta\gamma\,,
\ee
which holds for all triple of real numbers $a$, $b$ and $c$, and
$\alpha = \e^{-2a}$, $\beta = \e^{-2b}$, $\gamma = \e^{-2c}$.\\
This gives
\bea
\th\Jst>-\th\Js\th\Jt &\Leftrightarrow& \e^{-2\Jst}(\e^{-2\Js}+\e^{-2\Jt})<1+\e^{-2(\Js+\Jt)}
\nnm\\
&\Leftrightarrow& S+T<1+L\nnm\\
&\Leftrightarrow& l(1-s)(1-t)>-(1-s)(1-t)\,.\nnm
\eea
{\bf 4.} We now prove that \req{DualityCC} is one-to-one. It is sufficient to show that for
any triple $(\Js,\Jt,\Jst) \in \cD$ we can define a triple $(s,t,l)\in (]-1,1[)^3$ (see \req
{stl}) and that the corresponding triple $(\Js^*,\Jt^*,\Jst^*) \in \cD$.\\
We claim that $(s,t,l)$ is given by 
\bea
s = (1+S^2-T^2-L^2 - [(1+S^2-T^2-L^2)^2 -
4(S-TL)^2]^{\frac{1}{2}})/(2(S-TL))\,,\nnm\\
t = (1+T^2-S^2-L^2 - [(1+T^2-S^2-L^2)^2 -
4(T-SL)^2]^{\frac{1}{2}})/(2(T-SL))\,,\nnm\\
l = (1+L^2-S^2-T^2 - [(1+L^2-S^2-T^2)^2 -
4(L-ST)^2]^{\frac{1}{2}})/(2(L-ST))\,.\nnm\\\label{stl_STL}
\eea
{\bf 4.a} Let us verify that the quantities inside the square brackets are positive.\\
{\bf 4.a.1} $(1+L^2-S^2-T^2)^2 \geq 4(L-ST)^2$\\
We have
\bea\label{equ_d_1}
\th \Jst > -\th\Js\th\Jt &\Leftrightarrow& \e^{-2\Jst}(\e^{-2\Js}+\e^{-2\Jt})<1+\e^{-2(\Js+
\Jt)}\,.\nnm\\
&\Leftrightarrow& S+T<1+L\nnm\\
&\Leftrightarrow& S^2+T^2-1-L^2 < 2(L-ST)\nnm
\eea
where we have used \req{rel_th} and the fact that $S$, $T$, $L$ are positive.
Now if $\Jst\leq 0$ then
$$
L-ST =\e^{-2(\Js+\Jt)}(1-\e^{-4\Jst})\leq 0\,,
$$
hence,
$$
 (1+L^2-S^2-T^2)^2 > 4(L-ST)^2\,.
$$
On the other hand, if $\Jst>0$, we have
$$
\e^{-2\Jst}(\e^{-2\Jt}-\e^{-2\Js})<\e^{-2\Jt}-\e^{-2\Js}<1-\e^{-2(\Js+\Jt)}\,,
$$
which is equivalent to
$$
1-L>S-T \Leftrightarrow 1+L^2-S^2-T^2>2(L-ST)\,,
$$
where we have used the fact that $S\geq T$ if $\Js\geq\Jt$. This last expression finally gives
$$
(1+L^2-S^2-T^2)^2 \geq 4(L-ST)^2\,.
$$
{\bf 4.a.2} $(1+T^2-S^2-L^2)^2 \geq 4(T-SL)^2$\\
We have
$$
1+T^2-S^2-L^2=1+L^2-S^2-T^2 + 2(T^2-L^2) \geq 0\,,
$$
because $T^2-L^2 =\e^{-4\Js}(\e^{-4\Jst}-\e^{-4\Jt}) \geq 0$ if $\Jt\geq\Jst$.\\
On the other hand,
$$
\Jt>0 \Leftrightarrow T-SL\geq 0\,.
$$
Thus,
$$
(1+T^2-S^2-L^2)^2 \geq 4(T-SL)^2 \Leftrightarrow (T+1)^2 \geq (S-L)^2 \Leftrightarrow T+1 \geq
S-L\,,
$$
which holds if $\Js\geq \Jt$. Then use $1+L-S+T \geq S+T-S+T > 0$.\\ 
{\bf 4.a.3} $(1+S^2-T^2-L^2)^2 \geq 4(S-TL)^2$\\
Again $1+S^2-T^2-L^2 = 1+L^2-S^2-T^2 + 2(S^2-L^2) \geq 0$, and $S-TL>0$ if $\Js>0$.
So
$$
(1+S^2-T^2-L^2)^2 \geq 4(S-TL)^2 \Leftrightarrow S+1 \geq T-L\,,
$$
using the fact that $\Jst \leq \Jt$. The claim follows from $S+1+L-T\geq S+T+S-T> 0$.\\
{\bf 4.b} We now prove that $s$, $t$, $l \in ]-1,1[$.\\
{\bf 4.b.1} $s\geq 0$\\
As $S>LT$ (see 4.a.3.), it is enough to show that
$$
1+S^2-T^2-L^2 - [(1+S^2-T^2-L^2)^2 - 4(S-LT)^2]^\frac{1}{2} \geq 0\,,
$$
but this is obvious.\\
{\bf 4.b.2} $s < 1$\\
This is equivalent to show that
$$
4(S-LT)[2(S-LT)-(1+S^2-T^2-L^2)] < 0\,,
$$
which is a consequence of the above results (see 4.a.3.).\\
{\bf 4.b.3} $1>t\geq 0$\\
This is proved in the same way as for $s$.\\
{\bf 4.b.4} $\Jst\geq 0 \Rightarrow 1>l\geq 0$\\
$L-TS$ is positive and  we obtain the same kind of relations as for $s$.\\
{\bf 4.b.5} $\Jst< 0 \Rightarrow 0>l>-1$\\
This time we have  $L-TS<0$, which gives the results in the same way as before.\\
{\bf 4.c} It remains to show that $(\Js^*,\Jt^*,\Jst^*) \in \cD$.\\
We have already seen that $\Js^*>0$ (see 4.b.1.), $\Jt^*>0$ (see 4.b.3.), so we
just have to prove that $\Js^*\geq \Jt^*$,
$\Jt^*\geq\Jst^*$ and $\th\Jst^*>-\th\Js^*\th\Jt^*$.\\
{\bf 4.c.1} $\Js^*\geq \Jt^*$\\
\bea
\Js^*\geq \Jt^* &\Leftrightarrow& s\geq t\nnm\\
&\Leftrightarrow& \frac{(s-t)(1+l)}{1+stl}\geq 0\nnm\\
&\Leftrightarrow& S\geq T\nnm\\
&\Leftrightarrow& \Js \geq \Jt\,.\nnm
\eea
{\bf 4.c.2} $\Jt^*\geq\Jst^*$\\
In the same way,
$$
\Jt^*\geq\Jst^* \Leftrightarrow T\geq L \Leftrightarrow \Jt\geq\Jst\,.
$$
{\bf 4.c.3} $\th\Jst^*>-\th\Js^*\th\Jt^*$
$$
\th\Jst^*>-\th\Js^*\th\Jt^* \Leftrightarrow l> -st \Leftrightarrow L> 0\,.
$$
{\bf 4.d} The fact that \req{stl_STL} are solutions of \req{STL} can be checked by explicit
substitution.

\qed

\end{document}